\documentclass[amsmath,amssymb,twocolumn,eqsecnum]{revtex4}
\usepackage{graphicx}    
\usepackage[breaklinks, colorlinks, citecolor=blue]{hyperref}
\newcommand{\lied}[1]{{\rm L}_{#1}}

\newcommand{\virt}[0]{\hat{\delta}}

\newcommand{\half}[0]{\frac{1}{2}}
\newcommand{\cs}[3]{\Gamma^{#1}_{\,\,\,\, #2#3}}

\newcommand{\pd}[2]{\frac{\partial #1}{\partial #2}}
\newcommand{\ld}[0]{\mathcal{L}}

\newcommand{\dd}[0]{\textrm{d}}
\newcommand{\defn}[0]{\equiv}

\newcommand{\qsubrm}[2]{{#1}_{\scriptscriptstyle{\textrm{#2}}}}
\newcommand{\qsuprm}[2]{{#1}^{\scriptscriptstyle{\textrm{#2}}}}

\newcommand{\pis}[0]{ {\Pi}^{\scriptscriptstyle\rm{S}}}

\newcommand{\sol}[0]{\ld_{\scriptscriptstyle\{2\}}}
\newcommand{\rol}[0]{\mathcal{R}_{\scriptscriptstyle\{2\}}}

\newcommand{\hct}[0]{\mathcal{H}}
\renewcommand{\figurename}{Figure}
\newcommand{\ep}[0]{{ {\delta}_{\scriptscriptstyle{\rm{E}}}}}
\newcommand{\lp}[0]{{ {\delta}_{\scriptscriptstyle{\rm{L}}}}}
\def\be{\begin{equation}}
\def\ee{\end{equation}}
\def\bea{\begin{eqnarray}}
\def\eea{\end{eqnarray}}
\def\bse{\begin{subequations}}
\def\ese{\end{subequations}}

\newcommand{\fref}[1]{{Figure \ref{#1}}}

\let\oldsqrt\sqrt
\def\sqrt{\mathpalette\DHLhksqrt}
\def\DHLhksqrt#1#2{%
\setbox0=\hbox{$#1\oldsqrt{#2\,}$}\dimen0=\ht0
\advance\dimen0-0.2\ht0
\setbox2=\hbox{\vrule height\ht0 depth -\dimen0}%
{\box0\lower0.4pt\box2}}

\begin{document}
\title{Material  models of dark energy}
\author{Jonathan A. Pearson}
\email{jonathan.pearson@durham.ac.uk}
\affiliation{Centre for Particle Theory, Department of Mathematical Sciences, Durham University, South Road, Durham, DH1 3LE, U.K.}
\date{\today}

\begin{abstract}
We   review and develop a new class of ``dark energy'' models, in which   the relativistic theory of solids is used to construct material models of dark energy. These are models which include the effects of a continuous medium with well defined physical properties at the level of linearized perturbations. The formalism is constructed   for a medium with arbitrary symmetry, and then specialised to isotropic media (which   will be the case of interest for the majority of cosmological applications). We develop the theory of relativistic isotropic viscoelastic media whilst keeping in mind that we ultimately  want to   observationally constrain the allowed properties of the material model. We do this by obtaining the viscoelastic equations of state for perturbations (the entropy and anisotropic stress), as well as identifying the consistent corner of the theory which has constant equation of state parameter $\dot{w}=0$. We also connect to the non-relativistic theory of solids, by identifying the two quadratic invariants that are needed to construct the energy-momentum tensor, namely the Rayleigh dissipation function and Lagrangian for perturbations.  Finally, we develop the notion that the viscoelastic behavior of the medium can be thought of as a non-minimally coupled massive gravity theory. This  also provides a tool-kit for constructing consistent generalizations of coupled dark energy theories.

\end{abstract}

\maketitle
\tableofcontents

\section{Introduction}

The discovery of apparent cosmic acceleration \cite{Perlmutter:1998np, Riess:1998cb, Astier:2012ba} has spawned   huge interest in constructing dark energy \cite{Copeland:2006wr, dakrenergy_amendola} and modified gravity \cite{Durrer:2008in, Clifton:2011jh} theories capable of describing these observations.  There are many scalar field \cite{Tsujikawa:2006mw, PhysRevD.81.043520, DeFelice:2011th, Pujolas:2011he, Sawicki:2012re} and generalized scalar field \cite{Deffayet:2009mn, Goon:2011qf, DeFelice:2011hq, DeFelice:2011bh, Charmousis:2011bf, Charmousis:2011ea, Copeland:2012qf} models on the market, as well as a surge in the development of massive gravity \cite{ArkaniHamed:2002sp, Dubovsky:2005dw, Rubakov:2008nh, deRham:2010kj, Hinterbichler:2011tt, D'Amico:2011jj, Crisostomi:2012db, deRham:2014zqa} theories. The ultimate aim of these models is to provide some understanding of the underlying physical mechanisms causing the cosmic acceleration. There has also been considerable effort    going towards constructing ``model independent'' frameworks  \cite{Hu:1998kj, PhysRevD.69.083503, PhysRevD.76.104043, PhysRevD.77.103524, Skordis:2008vt, PhysRevD.81.083534, Hojjati:2011ix, PhysRevD.81.104023, Baker:2011jy, Zuntz:2011aq, Bloomfield:2011wa, Baker:2012zs, Battye:2012eu, Battye:2013aaa, Mueller:2012kb, Bloomfield:2012ff, Bloomfield:2013efa, Pourtsidou:2013nha, Battye:2013ida, Hu:2013twa, Cardona:2014iba} which can be used to confront classes of models or solutions of theories with observational data from, for example, WMAP \cite{Bennett:2012zja} and  Planck   \cite{Ade:2013ktc, Planck:2013kta, Ade:2013tyw}, as well as  galaxy weak lensing experiments \cite{Huterer:2001yu, Kilbinger:2012qz, Simpson:2012ra} and forecasting the potential discriminatory power of    experiments planned for the future  \cite{Abbott:2005bi, Laureijs:2011mu, Abate:2012za, Andre:2013afa}.

In this article we review and develop another  way to describe and understand the cause of cosmic acceleration. Rather than invoke the theory of scalar fields,  we will use the theory of relativistic solids to construct   \textit{material models of dark energy}.     The freedom in the theory of the solid corresponds to some physical property of the material (in contrast to the freedom in      a scalar field  theory, which corresponds to the kinetic or potential contributions to the  scalar field dynamics), which   can be     constrained using observations such as in the temperature and lensing anisotropies of the Cosmic Microwave Background (CMB) or the power spectrum of galaxy weak lensing.

 The theory of non-relativistic solids  \cite{ll_elast} is very well developed and has many diverse applications, whilst the theory of relativistic solids \cite{Synge:1959zz, 0305-4470-10-11-013, Palumbo1985, carter_springer_cond, Tartaglia:1995bx} is less  developed and has only a few applications, mainly in the description of neutron star crusts \cite{Carter21111972, BF01645505, Carter:1982xm, Carter08041991}. There is also a considerable body of literature pertaining to the construction of the theory of relativistic fluids, see e.g.,   \cite{Geroch:1990bw, Coley:1995uh, Maartens:1995wt, Maartens:1996vi, Maartens:1996dk, Maartens:1997sh, Maartens:1998xj, Hayward:1998hb, Chen:2001sma,  Muronga:2003ta,  Heinz:2005bw, Langlois:2006iq,   Romatschke:2009im, Azeyanagi:2009zd, Andersson:2006nr, Fukuma:2011pr, Ballesteros:2012kv, Andersson:2013jga, Ballesteros:2013nwa} and viscosity effects in cosmology \cite{Padmanabhan:1987dg, 0004-637X-506-2-485, Koivisto:2005mm, Mota:2007sz, Sapone:2013wda}. The   description of a relativistic perfectly elastic material model of dark energy has already been presented in the literature, for   isotropic \cite{Battye:1999eq, PhysRevD.60.043505, Battye:2005mm, PhysRevD.76.023005} and anisotropic  \cite{PhysRevD.74.041301, Battye:2009ze}  elastic solids. In addition, \cite{Boehmer:2013ss} present a modified gravity model using continuum mechanics, whilst \cite{Balek:2014uua, Balek:2014lza} study perfect elastic solids with positive pressure. A forthcoming paper will present the most up to date  constraints on the observationally allowed properties of the perfect elastic medium \cite{BattyePearsonMoss_edeconstraints}. The   focus of this   article is the development, and the application to cosmology, of the \textit{theory of relativistic viscoelastic solids}.

The material models of dark energy are not  best used as models which \textit{predict} the value of the equation of state parameter, $w = \qsubrm{P}{de}/\qsubrm{\rho}{de}$.  The material models have this parameter fixed by comparison to observations,   which forces other physical properties of the medium to adjust their values when computing, for example, the behavior of cosmological perturbations. As an example, consider perfectly elastic isotropic materials. There are two physical properties which characterise perturbations of the material: the bulk modulus, and shear (or, rigidity) modulus. For an elastic material,  only the bulk modulus is needed to construct $w$, but both the bulk and rigidity moduli are used to construct the sound speed (which is the only free parameter for linearized perturbations of the elastic medium). The perturbations are stable for a medium with negative pressure, i.e., $w<0$, but only if the rigidity is sufficiently large. By ``stable'' we mean that the sound speed is positive, and subliminal. The idea is that only certain ranges of values of these physical properties are allowed upon comparison to data.

The ``primary'' attractive feature of the material models is that there is a well prescribed rule-book for constructing the modified gravitational field equations that describe a material with a given physical property: the modified gravity field equations and corresponding free parameters gain physical interpretation.


One of the ``secondary''  interesting features of the material model is that the  evolution equation for the pressure of the solid is prescribed by the theory, after    time diffeomorphism invariance is imposed. This feature will become apparent in Section \ref{eq:sec:sub:components-tmunu}.

There are some advantages and drawbacks to both scalar field and material models of dark energy. First of all, a solid is much simpler to obtain a physical intuitive picture of than a scalar field. Also, one should recall that   only one scalar field has actually been observed in our Universe, but solids are common (to put it bluntly). Saying that, the mathematical   description of  a scalar field model is rather simple, compared to that needed to describe the material model. One of the attractive features of a material model is that it does not suffer from having to be carefully constructed to have a constant $w$ (it is actually quite simple and somewhat natural for a medium to have constant $w$), whereas scalar field models require substantial effort to do so.

The main ingredient of a model of a solid is   a constitutive relation between the stress tensor and the strain tensor. This constitutive relation then prescribes how the solid responds under  deformations. Stating what the pressure tensor is a function of is sufficient for isolating all freedom in the theory, and deducing how that freedom corresponds to physical properties of the material.

To account for  non-standard gravitational behavior (e.g., matter content which accelerates the Universe, or modified gravity), it is useful to append Einstein's gravitational field equations with a term on the right-hand-side,
\bea
G_{\mu\nu} = 8\pi G \left(T_{\mu\nu}+U_{\mu\nu}\right),
\eea
where $U_{\mu\nu}$ is the   \textit{dark energy-momentum tensor} which contains all contributions to the gravitational field equations due to whatever the physics is that is causing the apparent cosmic acceleration. Common examples include
\bea
U_{\mu\nu} \in \left\{ \begin{array}{cc} &(\mbox{scalar field})_{\mu\nu}\\&(\mbox{modified gravity})_{\mu\nu}\\ & (\mbox{material model})_{\mu\nu} \end{array} \right.,
\eea
where we have also included the material model concept in the list of possibilities. The main objective of this article is to understand what form $U_{\mu\nu}$ takes for material models.

Whilst the motivation for the current article comes from constructing a dark energy description, the theory applies equally well to other relativistic scenarios, and can be used, for example, in the context of inflation (see \cite{Endlich:2012pz, Bartolo:2013msa, Endlich:2013jia} and  \cite{Sitwell:2013kza}, where the latter paper used a formalism similar to ours).

In summary, the aim of this article is to describe dark energy via  the theory of solids; the ``physics'' of the material model we develop is
\begin{itemize}
\item {\textit{\textbf{Visco-elasticity}}} in which stress is a function of strain \textit{and} rate-of-strain.
\end{itemize}
The result of the article will be an understanding of how to include realistic modifications to a standard matter content of the Universe. The  novelty of these modifications is to include the effects of elastic and viscoelastic solids. We remain agnostic throughout as to whether these solids are supposed to be genuine solids, or a useful way to categorise the impact of more abstract modified gravity theories.

In the remainder of this introduction section we will recap the non-relativistic description of solids, which is mostly a review of Landau and Lifshitz \cite{ll_elast} and is included to aid the building of intuition. In Section \ref{sec:materialdescription} we build our material model of a viscoelastic medium, and in Section \ref{sec:eos-visco} we present  the viscoelastic equations of state for perturbations. In Section \ref{sec:timevariation} we discuss issues related to the time variation of the physical properties (such as $w$, the sound speeds, and dissipation coefficients), and in Section \ref{sec:connect} we point out a way of thinking about a viscoelastic medium in terms of more conventional types of dark energy/modified gravity theories. Final remarks and a summary of main results is given in Section \ref{sec:discussion}. The appendices hold some useful intermediate results and derivations. We collect some common symbols with their brief definitions and physical interpretation in Table \ref{tab:common}.


\begin{table*}
\begin{tabular}{||c |  c ||}
\hline
 Symbol & Meaning \\
 \hline
 $\lied{X}$ & Lie derivative operator along the vector $X^{\mu}$\\
 $u^{\mu}$ & Time-like unit vector \\
$\gamma_{\mu\nu} \defn g_{\mu\nu} + u_{\mu}u_{\nu}$ & Spatial metric, quantifying strain\\
$\lambda_{\mu\nu} \defn \lied{u}\gamma_{\mu\nu}$ & Rate-of-strain tensor\\
$K_{\mu\nu} = \nabla_{\mu}u_{\nu}$ &  Extrinsic curvature tensor\\
$\hct = \tfrac{1}{3}{K^{\mu}}_{\mu}$ & Hubble expansion\\
$T^{\mu\nu}$ & Energy-momentum tensor\\
$P^{\mu\nu}$ & Orthogonal pressure tensor\\
$w \defn \qsubrm{P}{de}/\qsubrm{\rho}{de}$ & Equation of state parameter\\
$\xi^{\mu} = (\chi, \xi^i)$ & Deformation vector\\
$\ep  $ & Eulerian perturbation: perturbation against background geometry \\
$\lp  \defn \ep + \lied{\xi}$ & Lagrangian variation: comoving with material medium\\
$E^{\mu\nu\alpha\beta} = E^{(\mu\nu)(\alpha\beta)}=E^{\alpha\beta\mu\nu}$ & Orthogonal elasticity tensor\\
$V^{\mu\nu\alpha\beta}=V^{(\mu\nu)(\alpha\beta)}$ & Orthogonal viscosity tensor\\
$w\Gamma$ & Entropy perturbation\\
$w\pis$ & Scalar anisotropic stress\\
$\{\beta, \lambda, \mu,\nu\}$ & Material properties\\
$\qsubrm{c}{s}^2, \,\,\qsubrm{c}{v}^2$ & Scalar and vector  sound  speeds\\
$\qsubrm{d}{s}, \,\,\qsubrm{d}{v}$ & Scalar and vector  damping coefficients\\
\hline
\end{tabular}\caption{Summary of commonly used symbols}\label{tab:common}
\end{table*}

\subsection{Hooke's law and Kelvin-Voigt solids}
Before we turn to the theory of relativistic solids, we shall review some important features from the theory  of non-relativistic solids. The idea is to give a relationship between the stress, $\sigma$, and strain, $\varepsilon$, due to   deformation  of a body (these are both rank-2 tensors, but for now we shall just consider the scalar relationship). The simplest example is a linear relationship between stress and strain,
\bea
\sigma = \beta \varepsilon.
\eea
This  is the defining characteristic of a \textit{Hookean solid}. The parameter $\beta$ is a property of the material, and dictates   the strength of the stress     from the given strain, and is related to the elastic modulus. The next simplest relationship is to include rate-of-strain
\bea
\sigma = \beta \varepsilon + \lambda \dot{\varepsilon}.
\eea
This is the defining characteristic of a \textit{Kelvin-Voigt solid}, which is a solid with   elastic and viscous behavior. The material properties are    the elastic modulus $\beta$, and  the coefficient of viscosity $\lambda$ (in the simple constitutive relation written above, these are both ``bulk'' moduli). 

The remainder of this paper is dedicated to rewriting these constitutive relations in increasing levels of sophistication, with the aim of constructing a material model which can be used to \textit{describe and interpret} the possible influences of viscoelastic dark energy. Before we jump to that we will carry on reviewing non-relativistic viscoelastic systems, paying particular attention to isotropic viscoelastic solids.

In a non-relativistic system, one should imagine that the coordinates of a medium in its relaxed state are given by $x^{i}$. A deformation alters these coordinates 
\bea
x^i\rightarrow x^i + \xi^i ,
\eea
where $\xi^i = \xi^i(x^j)$ is the \textit{deformation vector}. The strain tensor, $\varepsilon_{ij}$, is constructed  from symmetric combinations of the spatial derivatives of the deformation vector $\xi^{i}$ via
\bea
\label{eq:sec:non-rel-strain-lie-g}
\varepsilon_{ij} = \tfrac{1}{2} \big(\partial_i\xi_j + \partial_j \xi_i\big) = \partial_{(i}\xi_{j)}.
\eea
This is equivalent to setting the strain tensor to be the Lie derivative of the metric along the deformation vector, 
\bea
\varepsilon_{ij} = \tfrac{1}{2} \lied{\xi} g_{ij}.
\eea
The force, $F^i$, on a medium is computed by taking the divergence of  the stress tensor, $\sigma^{ij}$, 
\bea
F^i = \partial_j\sigma^{ij}.
\eea
The equation of motion of the deformations is constructed by relating the force due to stress, to the acceleration, $F^i = \rho\ddot{\xi}^i$, which yields
\bea
\label{eq:sec:eom-force-sigma}
\rho\ddot{\xi}^i = \partial_j\sigma^{ij}.
\eea
The task is to build a model which relates the stress tensor to the strain tensor: this will dictate the force on the body, and therefore the equation of motion of the deformations.

As we discussed above, the simplest model of a solid is embodied by Hooke's law, which relates the stress tensor to the strain tensor linearly. The most general way to do this (for a Hookean solid) is via a constitutive relation
\bea
\label{eq:sec:hookeslaw}
\sigma^{ij} = E^{ijkl}\varepsilon_{kl} ,
\eea
where $E^{ijkl}$ is the elasticity tensor (later on we will have much more to say about this tensor and its interpretation). To extend Hooke's law (\ref{eq:sec:hookeslaw}) we   construct the stress out of   more than just the strain, and the  easiest quantity to  introduce is the rate-of-strain  tensor, $\dot{\varepsilon}_{kl}$ (an overdot denotes derivative with respect to time). Hence, in the simplest extension, the stress $\sigma^{ij}$ is computed from the strain $\varepsilon_{ij}$ as
\bea
\label{eq:sec:kv-solid-defn-nonrel}
\sigma^{ij} = E^{ijkl}\varepsilon_{kl} +V^{ijkl}\dot{\varepsilon}_{kl}.
\eea
In addition to the elasticity tensor, we now have a viscosity tensor, $V^{ijkl}$. The relationship (\ref{eq:sec:kv-solid-defn-nonrel}) describes  a solid with elasticity and viscosity, known as a \textit{Kelvin-Voigt solid}. Using (\ref{eq:sec:non-rel-strain-lie-g}) to make the deformation vector explicit, the stress tensor of a viscoelastic solid is given by
\bea
\label{eq:sec:sigma-ef-visco-class}
\sigma^{ij} = E^{ijkl}\partial_{(k}\xi_{l)} +V^{ijkl} \partial_{(k}\dot{\xi}_{l)} ,
\eea
and   the equation of motion of the deformation vector (\ref{eq:sec:eom-force-sigma}) becomes
\bea
\label{eq:sec:eomclass}
\rho \ddot{\xi}^i  =  E^{ijkl}\partial_j\partial_{(k}\xi_{l)} + V^{ijkl}\partial_j \partial_{(k}\dot{\xi}_{l)},
\eea
where we took $E^{ijkl}$ and $V^{ijkl}$ to be constant throughout the medium.

The   elasticity and viscosity tensors, $E^{ijkl}$ and $V^{ijkl}$ respectively,  are what we call \textit{material tensors}.  The components of the material tensors are the physical properties of the medium, since they dictate how the medium   responds under strain.  The number of independent components of the material tensors are fixed by the symmetries of the medium. However, there are not as many components of the material tensors as there appears at first sight: there are some   symmetries in the indices, inherited from the fact that the stress and strain tensors are symmetric, and that the elasticity tensor can be derived from the elastic potential energy (we will have more to say about this in section \ref{eq:sec:intro-rayleigh}). These symmetries lead to the set of conditions
\bse
\label{eq:sec:symmet-class-ef}
\bea
\label{eq:sec:symmet-class-e}
E^{ijkl} = E^{(ij)(kl)} = E^{klij},
\eea
\bea
\label{eq:sec:symmet-class-f}
V^{ijkl} = V^{(ij)(kl)}.
\eea
\ese
The viscosity tensor gains an extra symmetry, namely the major symmetry under interchange of indices $V^{ijkl} = V^{klij}$, when the viscous theory is derived from a Rayleigh function (we have more to say about this in the next section). This extra symmetry is redundent for isotropic media.

For an isotropic medium, each of the material  tensors   have two free components. They are found by decomposing $E^{ijkl}$ and $V^{ijkl}$ into all possible combinations of the fundamental tensor (the metric, $\bar{g}_{ij}$) compatible with the symmetries (\ref{eq:sec:symmet-class-ef}),
\bse
\label{eq:class-decmmp-e-f}
\bea
\label{class-iso-elast}
E^{ijkl} &=&( \beta - \tfrac{2}{3}\mu)\bar{g}^{ij}\bar{g}^{kl} + 2 \mu \bar{g}^{i(k}\bar{g}^{l)j},\\
\label{class-iso-visc}
V^{ijkl} &=& ( \lambda - \tfrac{2}{3}\nu)\bar{g}^{ij}\bar{g}^{kl} + 2 \nu \bar{g}^{i(k}\bar{g}^{l)j}.
\eea
\ese
Physically, $\beta$ and $\mu$ are the bulk and shear elastic moduli respectively, and $\lambda$ and $\nu$ are the bulk and shear viscous moduli respectively: these are what we   call the physical \textit{material properties} of the solid. Using the  isotropic decompositions of the material tensors (\ref{eq:class-decmmp-e-f}),  the stress tensor (\ref{eq:sec:sigma-ef-visco-class}) becomes
\bea
\sigma^{ij} &=& (\beta - \tfrac{2}{3}\mu) \bar{g}^{ij} \partial_k\xi^k +(\lambda - \tfrac{2}{3}\nu) \bar{g}^{ij} \partial_k\dot{\xi}^k \nonumber\\
&&\qquad+ 2\mu\partial^{(i}\xi^{j)} + 2 \nu\partial^{(i}\dot{\xi}^{j)},
\eea
and the equation of motion for an isotropic viscoelastic medium (\ref{eq:sec:eomclass}) becomes
\bea
\label{eq:sec:class-eom-visco-isot}
\rho \ddot{\xi}^i &=& ( \beta + \tfrac{1}{3}\mu) \partial^i\partial_k\xi^k   +   \mu \partial_k\partial^k\xi^i  \nonumber\\
&& \qquad + ( \lambda + \tfrac{1}{3}\nu)\partial^i\partial_k\dot{\xi}^k+\nu \partial_k\partial^k\dot{\xi}^i.
\eea
Elastic waves have two sound speeds. This is simplest to see in the perfectly elastic case (by setting $\lambda = \nu=0$), and supposing that the deformations are a function of only one of the spatial coordinates: we shall take $\xi^i = \xi^i(t, x)$. Inserting this ansatz into  (\ref{eq:sec:class-eom-visco-isot}), the equations of motion governing each of  the three components of $\xi^i$ are
\bse
\bea
\pd{^2\xi^x}{t^2} -  {c}_{l}^2 \pd{^2\xi^x}{x^2}&=&0,\\
\pd{^2\xi^y}{t^2} -  {c}_{t}^2 \pd{^2\xi^y}{x^2}&=&0,\\
\pd{^2\xi^z}{t^2} -  {c}_{t}^2 \pd{^2\xi^z}{x^2}&=&0,
\eea
\ese
where
\bse
\label{eq:class-prop-speeds}
\bea
{c}_{l}^2 &\defn& \frac{\beta + \tfrac{4}{3}\mu}{\rho},\\
{c}_{t}^2 &\defn& \frac{\mu}{\rho}.
\eea
\ese
We now see that the elastic medium propagates two independent waves: there is a longitudinal wave travelling in the direction of the deformation with speed $c_l$, and a transverse wave travelling in a plane orthogonal to the deformation with speed $c_t$. The longitudinal wave is always faster than the transverse wave; infact, $c_l^2 \geq \tfrac{4}{3}c_t^2$. These two waves are also known as $P$ and $S$ waves respectively. The $S$-wave is only present due to the ability of the medium to support shear stresses.

 The elasticity theory we just discussed   was for non-relativistic media. There is a relativistic elasticity theory which was mostly constructed by Carter and collaborators. There are subtle complications for the extension to high-pressure relativistic media, since the ``strain'' tensor is now constructed out of an object which includes variations in the metric which leads to an  understanding of how metric fluctuations sources the deformation vector, and how the deformation vector sources the gravitational field equations.  

\subsection{Potential and Rayleigh functions}
\label{eq:sec:intro-rayleigh}
The theory of non-relativistic elastic solids can be derived from a quadratic elastic potential function, which is a   function of the strain.  The viscous part of the theory cannot be derived from any elastic potential. Instead, a second quadratic function needs to be introduced, and is called the \textit{Rayleigh  function}. Here we will briefly review how Rayleigh functions are used to construct non-relativistic system.

As a starting point,  consider the equation of motion of a damped pendulum:
\bea
m\ddot{x} = - kx - \alpha\dot{x}.
\eea
The first two terms can be derived from a Lagrangian, $L = \half m\dot{x}^2 - \half kx^2$, but the final  term which includes the effect of dissipation, cannot. The remedy is to introduce a second  ``master function'', in addition to the Lagrangian. This is known as  the Rayleigh   function, $R$. The Rayleigh function is introduced to enable the inclusion of velocity dependent contributions to   potential-like forces in the equation of motion. 

The dynamical system $D$ is constructed from the pair of invariants: $D = \{L, R\}$, and the equation of motion   is 
\bea
\label{eq:sec:eom-L-R}
\frac{\dd}{\dd t}\pd{L}{\dot{q_i}}  =\pd{L}{q_i}- \pd{R}{\dot{q}_i}.
\eea
Schematically, one imagines that the force contributions (i.e., the terms on the right-hand-side of the equation of motion) are respectively position and velocity dependent. The velocity dependent potentials would not be incorporated into a traditional Lagrangian theory. One should imagine that the equation of motion is given by
\bea
m\ddot{x}^i =\qsubrm{f}{tot}^i,\qquad \qsubrm{f}{tot}^i\defn \qsubrm{f}{pot}^i+\qsubrm{f}{dis}^i.
\eea
We decomposed the total force, $\qsubrm{f}{tot}^i$, into a term which comes from a potential, $\qsubrm{f}{pot}^i$, and a dissipative-force term, $\qsubrm{f}{dis}^i$,which is computed from the Rayleigh function $R$ via
\bea
\label{eq:sec:computing-fdiss}
\qsubrm{f}{dis}^i = - \pd{R}{\dot{q}_i}.
\eea

It is worth our providing an illustrative example. The Lagrangian which is quadratic in generalized coordinate velocities is
\bse
\label{eq:sec:quad-L-R-class}
\bea
L = \tfrac{1}{2} c^{ij} \dot{q}_i \dot{q}_j - V(q_i),
\eea
with $ c^{ij} = c^{ji}$. The  quadratic Rayleigh function is given by
\bea
R = \tfrac{1}{2}  d^{ij}\dot{q}_i\dot{q}_j ,
\eea
\ese
with $d^{ij} = d^{ji}$. The dissipative contribution to the force, using (\ref{eq:sec:computing-fdiss}), is $\qsubrm{f}{dis}^i = - d^{ij}\dot{q}_j$. Using the pair of invariants (\ref{eq:sec:quad-L-R-class}) to compute  the equation of motion (\ref{eq:sec:eom-L-R}) yields
\bea
c^{ij} \ddot{q}_j = - \pd{V}{q_i} - d^{ij}\dot{q}_j.
\eea

The appropriate potential and Rayleigh functions which give visocoelastic behavior are
\bse
\label{eq:sec:VR_non-rela-viso}
\bea
\label{eq:sec:VR_non-rela-viso-a}
U &=& \tfrac{1}{2}E^{ijkl}\varepsilon_{ij}\varepsilon_{kl},\\
\label{eq:sec:VR_non-rela-viso-b}
R &=& \tfrac{1}{2}V^{ijkl}\dot{\varepsilon}_{ij}\dot{\varepsilon}_{kl}.
\eea
\ese
The stress tensor is the sum of the derivatives of the potential with respect to the strain tensor and of the Rayleigh function with respect to the rate-of-strain tensor, 
\bea
\sigma^{ij} &=& \pd{U}{\varepsilon_{ij}}+\pd{R}{\dot{\varepsilon}_{ij}}
\eea
and so using (\ref{eq:sec:VR_non-rela-viso}), yields
\bea
\sigma^{ij}= E^{ijkl}\varepsilon_{kl}+V^{ijkl}\dot{\varepsilon}_{kl}.
\eea
We now see from (\ref{eq:sec:VR_non-rela-viso-b}) that requiring the viscoelastic theory to come from the pair of functions $\{U,R\}$ means that the elasticity and viscosity tensors automatically come endowed with  the set of symmetries
\bse
\bea
E^{ijkl} = E^{(ij)(kl)} = E^{klij},\\
V^{ijkl} = V^{(ij)(kl)} = V^{klij}.
\eea
\ese
%
%
%
 
\section{Material description}
\label{sec:materialdescription}
In this section we construct our relativistic viscoelastic theory from the formalism outlined by Carter and collaborators in \cite{Carter21111972, BF01645505, Carter:1973zz, Carter:1982xm, Carter08041991, carter_springer_cond}. The idea is to use  a matter manifold which is orthogonal to flow lines in the (four dimensional) space-time manifold, and is nicely explained   in \cite{PhysRevD.76.023005, Sitwell:2013kza}. All material quantities live on the matter space. Prescribing what the material quantities are a function of is sufficient for constructing a theory for perturbations with well defined physical interpretation.

This construction is easiest to work with via a (3+1)  decomposition of space-time, writing the metric as
\bea
g_{\mu\nu} = \gamma_{\mu\nu} - u_{\mu}u_{\nu},
\eea
where $\gamma_{\mu\nu}$ and $u_{\mu}$ are subject to the orthogonality and normality conditions
\bea
u^{\mu}\gamma_{\mu\nu}=0,\qquad u^{\mu}u_{\mu}=-1.
\eea
An orthogonal tensor is one which has vanishing contractions on any of its indices with the time-like unit vector $u_{\mu}$.
The space-time covariant derivative of the time-like unit vector defines the orthogonal extrinsic curvature tensor,
\bea
K_{\mu\nu} \defn \nabla_{\mu}u_{\nu},
\eea
with the following properties:
\bea
K_{\mu\nu} = K_{(\mu\nu)},\qquad u^{\mu}K_{\mu\nu}=0.
\eea
It is  useful to note that the extrinsic curvature is given by the Lie derivative of $\gamma_{\mu\nu}$ along the time-like vector $u^{\mu}$,
\bea
\label{eq:sec:ext-curv0liedgamma}
K_{\mu\nu} = \tfrac{1}{2}\lied{u}\gamma_{\mu\nu}.
\eea
Under deformation, the coordinates of the material udergo displacements $x^{\mu}\rightarrow x^{\mu} + \xi^{\mu}(x^{\nu})$. Under this deformation, the perturbation operator $\delta$ deforms as $\delta\rightarrow \delta + \lied{\xi}$, where $\lied{\xi}$ is the Lie derivative operator in the direction defined by the material deformation vector $\xi^{\mu}$. Two perturbation operators are now defined; $\ep$ is the perturbation with respect to some background space-time geometry, and $\lp$ is the perturbation which comoves with the deforming medium. These operators are related via
\bea
\lp = \ep + \lied{\xi}.
\eea
Respectively, these are Lagrangian and Eulerian variations.
For example, the metric perturbation which comoves with deformations of the medium, $\lp g_{\mu\nu}$ is given in terms of the metric perturbation with respect to a background space-time geometry, $\ep g_{\mu\nu}$, via
\bea
\label{eq:Sec:lp-g-e-g-xi}
\lp g_{\mu\nu} = \ep g_{\mu\nu} + 2 \nabla_{(\mu}\xi_{\nu)},
\eea
since $\lied{\xi}g_{\mu\nu} = 2\nabla_{(\mu}\xi_{\nu)}$. 

What remains to be presented is  the variation of the (orthogonal) pressure tensor, and consequently the variation of the energy-momentum tensor which sources the gravitational field equations. That requires a statement to be made about the ``physics'' of the medium: \textit{we will take the pressure tensor to be a function of  strain and rate-of-strain}. This is the defining characteristic of a viscoelastic medium.  Other choices are possible for the dependancies of the pressure tensor: we have picked this choice out of systematic simplicity. It is this choice which can be altered to change the physics of the medium.

 The result, which are the sources to the perturbed gravitational field equations due to an isotropic viscoelastic medium, is given by  (\ref{eq:sec:pert-compts-ep-visco}). But before that result, we show how to derive the variation for a general medium.

\subsection{Variation of the pressure and energy-momentum tensors}
The energy-momentum tensor of the medium is given by
\bea
\label{eq:sec:emt-proto-start}
T^{\mu\nu} = \rho u^{\mu}u^{\nu} + P^{ \mu\nu},
\eea
where $\rho$ is the energy density of the medium, and $P^{\mu\nu}$ is the orthogonal pressure tensor.
Without ambiguity,     (\ref{eq:sec:emt-proto-start}) can be varied to find
\bea
\label{eq:sec:lp_T-pre}
\lp T^{\mu\nu} =u^{\mu}u^{\nu}\lp \rho+ \lp P^{\mu\nu} +  \rho u^{\mu}u^{\nu}u^{\alpha}u^{\beta} \lp g_{\alpha\beta} ,
\eea
where we used the expression (\ref{eq:sec:appenx-id-1}) from the appendix for $\lp u^{\mu}$.
It is clear that we need input: we need to know expressions for $\lp \rho$ and $\lp P^{\mu\nu}$. The first is given from the conservation equation, as we now show. The second (as one can imagine) is what we shall use the machinery developed in appendix \ref{sec:-ofsarof} for;  most of what we collected and develop in the appendix are useful identities and relationships for orthogonal tensors.

The energy-momentum tensor (\ref{eq:sec:emt-proto-start}) is constrained by the conservation equation, $\nabla_{\mu}T^{\mu\nu} = 0$; using (\ref{eq:sec:emt-proto-start}) this yields
\bea
\big( \dot{\rho} + [\rho \gamma^{\alpha\beta} + P^{\alpha\beta}]K_{\alpha\beta}\big) u^{\nu} + {\gamma^{\mu}}_{\alpha}{\gamma^{\nu}}_{\beta}\nabla_{\mu}P^{\alpha\beta}=0.\nonumber\\
\eea
And so, demanding that $u_{\nu}\nabla_{\mu}T^{\mu\nu}=0$ yields
\bea
\label{eq:sec:cons-eq-one-bg}
\dot{\rho} = - [\rho \gamma^{\alpha\beta} + P^{\alpha\beta}]K_{\alpha\beta}.
\eea
This is recognisable as the usual fluid equation, albeit written in terms of the extrinsic curvature.
By direct calculation one can compute the Lie derivative along the time-like unit vector $u^{\mu}$ of the density $\rho$ and orthogonal metric $\gamma_{\mu\nu}$,
\bea
\lied{u}\rho = \dot{\rho},\qquad \lied{u}\gamma_{\mu\nu} = 2K_{\mu\nu},
\eea
so that (\ref{eq:sec:cons-eq-one-bg})  can be rephrased as
\bea
\label{eq:sec:cons-equ-in-lied}
\lied{u}\rho = - \tfrac{1}{2}\big(\rho \gamma^{\alpha\beta} + P^{\alpha\beta}\big) \lied{u}\gamma_{\alpha\beta}.
\eea
Replacing the Lie derivative with Lagrangian variation in  (\ref{eq:sec:cons-equ-in-lied}) yields
\bea
\label{eq:sec;lp-rho}
\lp \rho =- \tfrac{1}{2}(\rho \gamma^{\alpha\beta} + P^{\alpha\beta}) \lp g_{\alpha\beta},
\eea
where we note that the prefactor of $ \lied{u}\gamma_{\alpha\beta}$ in (\ref{eq:sec:cons-equ-in-lied}) is orthogonal.   Therefore, using (\ref{eq:sec;lp-rho}),  (\ref{eq:sec:lp_T-pre}) becomes
\bea
\label{eq:sec:lpt-pre-Plp}
\lp T^{\mu\nu} &=& \lp P^{\mu\nu} - \tfrac{1}{2} \big[ u^{\mu}u^{\nu}P^{\alpha\beta} + \rho u^{\mu}u^{\nu} \gamma^{\alpha\beta}\nonumber\\
&& - 2 \rho u^{\mu}u^{\nu}u^{\alpha}u^{\beta}\big] \lp g_{\alpha\beta}.
\eea
This is a general expression, and all we need now is $\lp P^{\mu\nu}$. 

The important point we now need to come back to and utlitise  is that the pressure tensor, $P^{\mu\nu}$, is an orthogonal tensor function of strain and rate-of-strain: here we will draw together a lot of the machinery which we laid out in    appendix \ref{sec:-ofsarof} in order to compute its variation. From that, we will compute the variation of the energy-momentum tensor for a viscoelastic solid.

The   spatial metric, $\gamma_{\mu\nu}$, quantifies the strain of the medium. The rate-of-strain tensor, $\lambda_{\mu\nu}$, is the Lie derivative of the strain tensor in the time-like direction,
\bea
\lambda_{ \mu\nu} \defn \lied{u}\gamma_{ \mu\nu}  .
\eea
When the pressure is a function of strain, $\gamma_{\mu\nu}$, and rate-of-strain, $\lambda_{\mu\nu}$, it follows that its variation is given by
\bea
\label{eq:sec:vary-P-cov-1}
\lp P_{\mu\nu} = \pd{P_{\mu\nu}}{\gamma_{\alpha\beta}} \lp \gamma_{\alpha\beta} + \pd{P_{\mu\nu}}{\lambda_{\alpha\beta}} \lp \lambda_{\alpha\beta}.
\eea
We remind that $\lambda_{\mu\nu}$ is  related to the extrinsic curvature tensor via (\ref{eq:sec:ext-curv0liedgamma}).
By (\ref{eq:sec:lp-P-fljdhjkfhd}), (\ref{eq:sec:vary-P-cov-1}) can be written as
\bea
\label{eq:sec:vary-P-cov}
\lp P_{\mu\nu} = \pd{P_{\mu\nu}}{g_{\alpha\beta}} \lp g_{\alpha\beta} + \pd{P_{\mu\nu}}{K_{\alpha\beta}} \lp K_{\alpha\beta}.
\eea
Carefully raising indices on the left-hand side of (\ref{eq:sec:vary-P-cov}) by using (\ref{eq:sec;change-upp-todwnn}), yields
\bea
\label{eq:lp-P-visc}
\lp P^{\mu\nu} &=& - \tfrac{1}{2} \big[ E^{\mu\nu\alpha\beta} + P^{\mu\nu}\gamma^{\alpha\beta} - 4 P^{\alpha(\mu}u^{\nu)}u^{\beta}  \big] \lp g_{\alpha\beta} \nonumber\\
&& -   V^{\mu\nu\alpha\beta}\lp K_{\alpha\beta},
\eea
where we   defined the derivatives of the pressure tensor with respect to strain and rate-of-strain as
\bse
\label{eq:sec:defn-ortho-elas-visc}
\bea
\pd{P^{\rho\sigma}}{\gamma_{\alpha\beta}} &\defn& - \tfrac{1}{2} (E^{\rho\sigma \alpha\beta} + P^{\rho\sigma}\gamma^{ \alpha\beta}),
\eea
\bea
 \pd{P^{\rho\sigma}}{K_{\alpha\beta}}&\defn& -  V^{\rho\sigma \alpha\beta}.
\eea
\ese
The first three terms in  (\ref{eq:lp-P-visc}) are exactly those present for a perfect elastic solid. The last one is due to the fact that the system is a function of the rate-of-strain in addition to the strain.  This is the viscous contribution.

Using our derived expression for $\lp P^{\mu\nu}$ (\ref{eq:lp-P-visc})  in the general expression for $\lp T^{\mu\nu}$ (\ref{eq:sec:lpt-pre-Plp})   yields
\bea
\label{eq:sec:lp-T}
\lp T^{\mu\nu}  &=& - \tfrac{1}{2} \big[ W^{\mu\nu\alpha\beta}  + T^{\mu\nu}g^{\alpha\beta} \big] \lp g_{\alpha\beta} -   V^{\mu\nu\alpha\beta} \lp K_{\alpha\beta},\nonumber\\
\eea
in which we  defined a non-orthogonal ``elasticity tensor'', for convenience, as 
\bea
\label{eq:sec:gen-elast0tensor-fdkjfhd}
W^{\mu\nu\alpha\beta} &\defn& E^{\mu\nu\alpha\beta} + P^{\mu\nu}u^{\alpha}u^{\beta} + P^{\alpha\beta} u^{\mu}u^{\nu}\nonumber\\
&&  - 4u^{(\alpha}P^{\beta)(\mu}u^{\nu)} - \rho u^{\mu}u^{\nu}u^{\alpha}u^{\beta}.
\eea
Equation (\ref{eq:sec:lp-T}) is the first of our main results. This expression acts as the source to the perturbed gravitational field equations for a viscoelastic medium, since     the pressure tensor is a function of strain and rate of strain. 

Respectively, $E^{\rho\sigma\mu\nu}$ and $V^{\rho\sigma\mu\nu}$ are the elasticity  and viscosity tensors: they are the material tensors, and their components contain all material properties of the medium which is being described. The material tensors  have the following symmetries in their indicies:
\bse
\label{eq:sec:symm-e0f}
\bea
E^{\rho\sigma\mu\nu}  = E^{(\rho\sigma)(\mu\nu)}  = E^{\mu\nu\rho\sigma} ,
\eea
\bea
 V^{\rho\sigma\mu\nu} = V^{(\rho\sigma)(\mu\nu)}.
\eea
\ese
The major symmetry of index interchange of the elasticity tensor is due to the fact that the elasticity tensor is related to the elastic potential energy, which is the coefficient of quadratic combinations of the strain tensor. From (\ref{eq:sec:defn-ortho-elas-visc}), it is apparent that the material tensors  are orthogonal
\bse
\bea
u_{\mu}E^{\rho\sigma\mu\nu} =0,
\eea
\bea
 u_{\mu} V^{\rho\sigma\mu\nu} =u_{\rho} V^{\rho\sigma\mu\nu} =0.
\eea
\ese

We now   show how to compute the components of the mixed Eulerian perturbed energy-momentum tensor, $\ep {T^{\mu}}_{\nu}$: these are the sources to the perturbed gravitational field equations. The contravariant components of the Eulerian perturbed energy-momentum tensor are given in terms of the Lagrangian perturbed components by
\bea
\ep T^{\mu\nu} = \lp T^{\mu\nu} - \lied{\xi}T^{\mu\nu},
\eea
and so the   components of the mixed Eulerian perturbed energy-momentum tensor are given by
\bea
\label{eq:sec:ep-t-from-lp-t-}
\ep {T^{\mu}}_{\nu} = g_{\nu\alpha}\lp T^{\mu\alpha} - g_{\nu\alpha}\lied{\xi}T^{\mu\alpha} + T^{\mu\alpha}\ep g_{\nu\alpha},
\eea
where we remind that 
\bea
\label{eq:Sec:lied-T}
\lied{\xi}T^{\mu\nu} = \xi^{\alpha}\nabla_{\alpha}T^{\mu\nu} - 2 T^{\alpha(\mu}\nabla_{\alpha}\xi^{\nu)}.
\eea
The Lagrangian variation of the metric is given by (\ref{eq:Sec:lp-g-e-g-xi}),  and the Lagrangian variation of the extrinsic curvature tensor is  given by 
\bea
\label{eq:sec:lp-k-withxiexpliciy}
\lp K_{\mu\nu}  &=& \ep K_{\mu\nu}+2u^{(\alpha} {\gamma^{\beta)}}_{(\mu}\nabla_{\nu)} \nabla_{\alpha}\xi_{\beta}  \nonumber\\
&&-  u^{\alpha}u^{\beta}u_{(\mu}\nabla_{\nu)} \nabla_{\alpha}\xi_{\beta}-  u^{\alpha}\nabla_{(\mu}\nabla_{\nu)}\xi_{\alpha} \nonumber\\
&&  + 2\big[({\gamma^{\alpha}}_{(\mu} - \tfrac{1}{2}u_{(\mu}u^{\alpha}){K^{\beta}}_{\nu)} + u^{\beta}{K^{\alpha}}_{(\mu}u_{\nu)}\big]\nabla_{(\alpha}\xi_{\beta)}   \nonumber\\
&&- u_{\alpha}{R^{\alpha}}_{(\mu\nu)\beta}\xi^{\beta}.
\eea
The derivation of (\ref{eq:sec:lp-k-withxiexpliciy}) is given   in appendix \ref{sec:var-extr-curv}.
Putting  (\ref{eq:sec:lp-T}) and (\ref{eq:Sec:lied-T}) into (\ref{eq:sec:ep-t-from-lp-t-}), and using (\ref{eq:Sec:lp-g-e-g-xi}) and (\ref{eq:sec:lp-k-withxiexpliciy}) to replace the remaining Lagrangian variations, we then obtain
\bea
\label{eq:sec:ep-t-mixed-cov}
\ep {T^{\sigma}}_{\lambda} &=& - \tfrac{1}{2}({W^{\sigma}}_{\lambda}{}^{\mu\nu} + {T^{\sigma}}_{\lambda}g^{\mu\nu})\ep g_{\mu\nu} - {V^{\sigma}}_{\lambda}{}^{\mu\nu}\ep K_{\mu\nu} \nonumber\\
&&+ T^{\sigma\alpha}\ep g_{\alpha\lambda} - {V^{\sigma}}_{\lambda}{}^{\mu\nu}[2u^{(\alpha} {\gamma^{\beta)}}_{(\mu}\nabla_{\nu)} \nabla_{\alpha}\xi_{\beta}\nonumber\\
&&  -  u^{\alpha}\nabla_{(\mu}\nabla_{\nu)}\xi_{\alpha}   ]-[{W^{\sigma}}_{\lambda}{}^{ \alpha\beta} + 2{V^{\sigma}}_{\lambda}{}^{\mu\nu}{\gamma^{(\alpha}}_{(\mu} {K^{\beta)}}_{\nu)} \nonumber\\
&&+ {T^{\sigma}}_{\lambda}g^{ \alpha\beta}    -2 T^{\alpha(\sigma}{g^{\beta}}_{\lambda)}]\nabla_{\alpha}\xi_{\beta}\nonumber\\
&&+ [{V^{\sigma}}_{\lambda}{}^{\mu\nu}u_{\alpha}{R^{\alpha}}_{(\mu\nu)\beta}- \nabla_{\beta}{T^{\sigma}}_{\lambda}]\xi^{\beta} .
\eea
We used the orthogonality of the viscosity tensor, $V^{\mu\nu\alpha\beta}$. A consequence of this orthogonality is that there are no viscous contributions to the time-time and time-space projections of $\ep {T^{\sigma}}_{\lambda}$. We will give explicit expressions for the components later on.

The energy-momentum tensor satisifies the perturbed conservation equation,
\bea
\label{eq:sec:eom-t-ep}
\ep (\nabla_{\mu}{T^{\mu}}_{\nu})=0.
\eea
This also acts like the equation of motion.
For any energy-momentum tensor, (\ref{eq:sec:eom-t-ep}) yields
\bea
&&g_{\nu\alpha}\nabla_{\mu}\ep T^{\mu\alpha} + T^{\mu\alpha}\nabla_{\mu} \ep g_{\nu\alpha}\nonumber\\
&& \qquad+ {T^{\alpha}}_{\nu}\ep \cs{\mu}{\mu}{\alpha} - {T^{\mu}}_{\alpha}\ep \cs{\alpha}{\mu}{\nu}=0.
\eea
Since $\ep T^{\mu\nu} = \lp T^{\mu\nu} - \lied{\xi}T^{\mu\nu}$, this becomes
\bea
&&2g_{\nu\alpha}T^{\beta(\mu}\nabla_{\mu}\nabla_{\beta}\xi^{\alpha)}-  (\nabla_{\beta}{T^{\mu}}_{\nu})\nabla_{\mu}\xi^{\beta} +g_{\nu\alpha}\nabla_{\mu}\lp T^{\mu\alpha} \nonumber\\
&& \qquad= 2{T^{\mu}}_{[\alpha}\ep \cs{\alpha}{\nu]}{\mu} -T^{\mu\alpha}\nabla_{\mu} \ep g_{\nu\alpha} . 
\eea
The derived perturbed energy-momentum tensor (\ref{eq:sec:lp-T}) can then be inserted.  The resulting expression is highly convoluted to write down explicitly, but we will do so for an isotropic medium in the next section.
 
\subsection{Isotropic medium}
So far we have assumed nothing about the symmetries of the medium. From now on we shall take the medium to be spatially isotropic and homogeneous. This assumption about the symmetry of the medium is not nessecary for previous results to hold, for example (\ref{eq:sec:ep-t-mixed-cov}). However, without the assumption, the resulting equations become highly unweildy. That said: since we are constucting a model of a medium with an application in cosmology in mind, assuming it to be isotropic is quite a sensible restriction (that said, the case of a cosmological anisotropic perfect elastic medium was studied in detail in \cite{Battye:2006mb, Battye:2009ze}).

The assumption of isotropy is implemented   in the decomposition of the pressure and material tensors. These tensors are decomposed into the most fundamental isotropic tensor, which is also orthogonal. The only such tensor is the spatial metric, $\gamma_{\mu\nu}$. What this means is that  the pressure tensor for an isotropic medium is given in terms of the pressure scalar $P$ as
\bea
P^{\mu\nu} = P\gamma^{\mu\nu}.
\eea
The decomposition of the material tensors, $E^{\mu\nu\alpha\beta}$ and $V^{\mu\nu\alpha\beta}$, has a little more freedom. Given the assumption of isotropy and the symmetries in their indicies (\ref{eq:sec:symm-e0f}), the material tensors completely decompose as
\bse
\label{eq:sec:decomp-iso-materialtensprs}
\bea
E^{\mu\nu\alpha\beta} &=& (\beta-P-\tfrac{2}{3}\mu) \gamma^{\mu\nu}\gamma^{\alpha\beta} + 2 (\mu+P)\gamma^{\mu(\alpha}\gamma^{\beta)\nu},\nonumber\\ \\
V^{\mu\nu\alpha\beta} &=& a(\lambda-\tfrac{2}{3}\nu)\gamma^{\mu\nu}\gamma^{\alpha\beta} + 2 a\nu\gamma^{\mu(\alpha}\gamma^{\beta)\nu}.
\eea
\ese
There are four pieces of freedom here: $\{\beta,\mu,\lambda,\nu\}$. These are the material properties, and are dimensionful; later on we will obtain the dimensionless freedom in the theory. 
The calculations we are about to perform will concrete their physical interpretation, but for now the meaning of these pieces of freedom are:
\bea
\underbrace{\left. \begin{array}{c}\beta :  \mbox{bulk  }\\
\mu  :   \mbox{shear  }
\end{array}\right.}_{\mbox{elastic moduli}},\qquad
\underbrace{\left. \begin{array}{c}\lambda :  \mbox{bulk}\\
\nu  :   \mbox{shear}
\end{array}\right.}_{\mbox{viscous moduli}}.
\eea

\subsubsection{Components of $\ep {T^{\mu}}_{\nu}$}
\label{eq:sec:sub:components-tmunu}
We will  compute the components of the perturbed energy-momentum tensor which sources gravitational field perturbations
\bea
\ep {G^{\mu}}_{\nu} = 8 \pi G \ep {T^{\mu}}_{\nu} 
\eea
for an isotropic viscoelastic medium. Recall that we presented the covariant form of the components $\ep {T^{\mu}}_{\nu} $ in (\ref{eq:sec:ep-t-mixed-cov}).

We compute in the synchronous gauge, on a conformally flat FRW background; this means that we set $\ep g_{\mu\nu} = a^2(\tau)h_{\mu\nu}$ with $h_{00}=h_{0i}=0$, and overdots will denote derivatives with respect to conformal time $\tau$. In particular, the Hubble expansion is defined via
\bea
\hct = \tfrac{1}{3}{K^{\mu}}_{\mu} = \dot{a}/a.
\eea
The components of the deformation field are $\xi^{\mu}= (\chi, \xi^i)$, where $u_{\mu}\xi^{\mu}=\chi$. Even though we are working in the synchronous gauge, our results will turn out to be gauge invariant.

For the components of the Lagrangian perturbed metric (\ref{eq:Sec:lp-g-e-g-xi}) we find
\bse
\bea
\lp g_{00} &=& - 2 a^2(\dot{\chi} + \hct\chi),
\\
 \lp g_{0i} &=& a^2(\xi_i - \partial_i\chi),
\\
\lp g_{ij} &=& a^2(h_{ij} + 2 \partial_{(i}\xi_{j)} + 2 \hct\chi \delta_{ij}).
\eea
\ese
For the components of the perturbed extrinsic curvature (\ref{eq:sec:lp-k-withxiexpliciy}) we find
\bse
\label{eq:sec:comps-lagper-extcurv-syn}
\bea
\lp K_{00} &=& 0,
\\
\lp K_{0i} &=& \tfrac{1}{2} a\big[ \ddot{\xi}_i + \hct \dot{\xi}_i\big],\\
\label{eq:sec:comps-lagper-extcurv-syn-c}
\lp K_{ij} &=& \tfrac{1}{2} a\big[ \dot{h}_{ij} + 2\partial_{(i}\dot{\xi}_{j)} + 2 \hct(h_{ij} + 2 \partial_{(i}\xi_{j)} )\big] + \ddot{a}\chi\delta_{ij}.\nonumber\\
\eea
\ese
Notice the existence of the $\ddot{a}\chi\delta_{ij}$-term in (\ref{eq:sec:comps-lagper-extcurv-syn-c}); this will have some interesting consequences.

The   components of $\ep {T^{\mu}}_{\nu}$ are computed from (\ref{eq:sec:ep-t-mixed-cov}), using the  isotropic decompositions of the material tensors given in (\ref{eq:sec:decomp-iso-materialtensprs})  and yield
\bse
\label{eq:sec;comps-pre-chiremoval}
\bea
\ep {T^0}_0 = [\dot{\rho}+3\hct(\rho+P)]\chi +   (\rho+P)\bigg(\tfrac{1}{2} h + \partial_i\xi^i\bigg),\nonumber\\
\eea
\bea
\ep {T^i}_0 = -(\rho+P)\dot{\xi}^i,
\eea
\bea
\ep {T^i}_j &=&-  (\beta-\tfrac{2}{3}\mu) \bigg(\tfrac{1}{2}h+\partial_k\xi^k\bigg) {\delta^i}_j-\mu\bigg({h^i}_j+2\partial^{(i}\xi_{j)}\bigg)\nonumber\\
&&-(\lambda-\tfrac{2}{3}\nu)\bigg(\tfrac{1}{2}\dot{h} + \partial_k\dot{\xi}^k +2\hct[\tfrac{1}{2}h+\partial_k\xi^k]\bigg) {\delta^i}_j\nonumber\\
&&-\nu \bigg( \dot{h}{{}^i}_j+2 \partial^{(i}\dot{\xi}_{j)}+2 \hct[{ {h}^i}_j+2\partial^{(i}\xi_{j)}] \bigg)\nonumber\\
&& -\bigg( \dot{P}+ 3\beta\hct+3\lambda\tfrac{ \ddot{a} }{a}  \bigg) \chi{\delta^i}_j  .
\eea
\ese

The final thing we want to do is to obtain the conditions placed on (\ref{eq:sec;comps-pre-chiremoval}) which leave behind components of the Eulerian perturbed energy-momentum tensor which are invariant under time diffeomorphisms, but not spatial ones. In some sense, this is a highly desirable concept when designing a model of a solid:  intuitively, solids fluctutate in space, but not time. One does not need to impose this condition, but doing so enables highly desirable physical interpretation and some other very useful properties which will become apparent.

To get the desired conditions, we   imagine that off-foliation diffeomorphisms are allowed, so that $u_{\mu}\xi^{\mu}\neq 0$, but we want them to have no effect on the system. By inspecting the components (\ref{eq:sec;comps-pre-chiremoval}), to decouple $\chi = u_{\mu}\xi^{\mu}$, we require
\bse
\label{eq:conds-to-decoup_chi}
\bea
\dot{\rho} +3\hct(\rho+P)=0,
\eea
\bea
\label{eq:sec:evol-press-re-decoup_chi}
\dot{P}+ 3\beta\hct+3\lambda(\dot{\hct} + \hct^2)=0.
\eea
\ese
The first condition is just the continuity equation for the energy density, and the second condition imposes an evolution rule for the pressure. From the condition (\ref{eq:sec:evol-press-re-decoup_chi}), one can obtain
\bea
\label{eq:sec:dpdrho-fromdecoupchi}
(\rho+P)\frac{\dd P}{\dd\rho} = \beta + \lambda\frac{\dot{\hct} + \hct^2}{\hct}.
\eea

Applying the conditions (\ref{eq:conds-to-decoup_chi}) to the components (\ref{eq:sec;comps-pre-chiremoval}) yields
\bse
\label{eq:sec:pert-compts-ep-visco}
\bea
\ep {T^0}_0 =    (\rho+P)\bigg(\tfrac{1}{2} h + \partial_i\xi^i\bigg),
\eea
\bea
\ep {T^i}_0 = -(\rho+P)\dot{\xi}^i,
\eea
\bea
\label{eq:sec:pert-compts-ep-visco-space}
\ep {T^i}_j &=&-  (\beta-\tfrac{2}{3}\mu) \bigg(\tfrac{1}{2}h+\partial_k\xi^k\bigg) {\delta^i}_j\nonumber\\
&&-2\mu\bigg(\tfrac{1}{2}{h^i}_j+\partial^{(i}\xi_{j)}\bigg)\nonumber\\
&&-(\lambda-\tfrac{2}{3}\nu)\bigg(\tfrac{1}{2}\dot{h} + \partial_k\dot{\xi}^k +2\hct[\tfrac{1}{2}h+\partial_k\xi^k]\bigg) {\delta^i}_j\nonumber\\
&&-2\nu \bigg( \tfrac{1}{2}\dot{h}{{}^i}_j+  \partial^{(i}\dot{\xi}_{j)}+2 \hct[\tfrac{1}{2}{ {h}^i}_j+ \partial^{(i}\xi_{j)}] \bigg). \nonumber\\
\eea
\ese
The terms on the first two lines of $\ep {T^i}_j$ are the spatial parts of the strain tensor: there is a diagonal contribution, and an off-diagonal contribution. On the third and fourth lines we observe the spatial parts of the rate of strain tensor (again, with diagonal and off-diagonal contributions).  It is relatively obvious that unless $\lambda=0$, the perturbed pressure will not be proportional to the perturbed density -- this is a classic hall-mark of a non-adiabatic system which we will further elucidate  later on. Therefore, (\ref{eq:sec:pert-compts-ep-visco}) are the  expressions for fluctuations of a relativistic non-adiabatic  viscoelastic medium.  

We will be performing a suite of small calculations to build up intuition of terms in both the energy-momentum tensor, and the equations of motion. The first thing we want to point out is the connection between the expression for the perturbed pressure tensor of the relativistic system (\ref{eq:sec:pert-compts-ep-visco-space}) and the corresponding expression for a non-relativistic system, (\ref{eq:sec:kv-solid-defn-nonrel}). By defining ``stress'',  ``strain'' and ``rate-of-strain'' tensors,
\bse
\bea
{\sigma^i}_j\defn \ep{T^i}_j,
\eea
\bea
{\varepsilon^i}_j \defn \tfrac{1}{2}{h^i}_j + \partial^{(i}\xi_{j)},\qquad \widehat{\dot{\varepsilon}}{{}^i}_j \defn \dot{\varepsilon}{{}^i}_j + 2 \hct {\varepsilon^i}_j,
\eea
\ese
the spatial part (\ref{eq:sec:pert-compts-ep-visco-space}) can be written in a rather suggestive form:
\bea
\label{eq:Sec:emt-ij-withepsilon}
{\sigma^i}_j&=&-  \beta {\varepsilon^k}_k  {\delta^i}_j-2\mu({\varepsilon^i}_j-\tfrac{1}{3}{\varepsilon^k}_k  {\delta^i}_j)  \nonumber\\
&&-\lambda \widehat{\dot{\varepsilon}}{{}^k}_k {\delta^i}_j-2\nu(\widehat{\dot{\varepsilon}}{{}^i}_j -\tfrac{1}{3}\widehat{\dot{\varepsilon}}{{}^k}_k {\delta^i}_j).
\eea
It certainly looks like the stress tensor is constructed from the strain tensor and the rate of strain tensor, which was the defining characteristic of a viscoelastic medium; infact, of a Kelvin-Voigt solid.

The deformations of an isotropic medium come in two types: compression and shear. These are characterized by a strain tensor which is pure-diagonal and pure-off-diagonal respectively. For deformations which are purely of these types,   (\ref{eq:Sec:emt-ij-withepsilon}) becomes
\begin{itemize}
\item  Compression:
\bea
{\sigma^i}_j =-  \beta {\varepsilon^k}_k  {\delta^i}_j  -\lambda \widehat{\dot{\varepsilon}}{{}^k}_k {\delta^i}_j ,
\eea
\item  Shear:
\bea
{\sigma^i}_j= -2\mu({\varepsilon^i}_j-\tfrac{1}{3}{\varepsilon^k}_k  {\delta^i}_j)  -2\nu(\widehat{\dot{\varepsilon}}{{}^i}_j -\tfrac{1}{3}\widehat{\dot{\varepsilon}}{{}^k}_k {\delta^i}_j).
\eea
\end{itemize}

This enables us to read off the physical interpretation of the various free coefficients. Firstly, $\beta$ is the coefficient of elastic compression, which is also called the bulk modulus. Second, $\mu$ is the coefficient of elastic shear deformations,  the shear modulus. Third, $\lambda$ is the coefficient of viscous compression (viscous bulk modulus), and finally, $\nu$ is the coefficient of viscous shear deformations (viscous shear modulus).

\subsubsection{Equation of motion}
The equation of motion of the deformation vector is given by ${\gamma^{\nu}}_{\alpha}\ep(\nabla_{\mu}{T^{\mu}}_{\nu})=0$, where (\ref{eq:sec:pert-compts-ep-visco}) is used for the $\ep {T^{\mu}}_{\nu}$. This yields
\bse
\label{eq:Sec:eom-diff-full}
\bea
&& (\rho+P)[\ddot{\xi}^i+\hct\dot{\xi}^i] -3[\beta\hct + \lambda(\dot{\hct} + \hct^2) ]\dot{\xi}^i \nonumber\\
&&- (\lambda+\tfrac{1}{3}\nu)[\partial^i\partial_k\dot{\xi}^k+2\hct\partial^i\partial_k\xi^k]    -\nu[\partial_k\partial^k\dot{\xi}^i+2\hct\partial_k\partial^k\xi^i]\nonumber\\
&&  - (\beta+\tfrac{1}{3}\mu) \partial^i\partial_k\xi^k  - \mu\partial_k\partial^k\xi^i = S^i_{[h]}  ,
\eea
where we defined the source due to the metric perturbations, $S^i_{[h]}$, as
\bea
S^i_{[h]}&\defn& (\lambda-\tfrac{2}{3}\nu)\tfrac{1}{2}[ \partial^i\dot{h}+2\hct\partial^ih]  +\nu\big[\partial^j\dot{h}{{}^i}_j  +2\hct  \partial^j{h^i}_j\big]\nonumber\\
&&+(\beta-\tfrac{2}{3}\mu) \tfrac{1}{2} \partial^ih +\mu\partial^j{h^i}_j .
\eea
\ese
This should be compared with the non-relativistic equation of motion for a viscoelastic medium, (\ref{eq:sec:class-eom-visco-isot}).
\subsubsection{Scalar-vector-tensor split}
We will now perform a scalar-vector-tensor (SVT) split \cite{Bardeen:1980kt, Mukhanov1992203} of    the components of the Eulerian perturbed energy-momentum tensor, $\ep {T^{\mu}}_{\nu}$  (\ref{eq:sec:pert-compts-ep-visco}), and the equation of motion of the deformation vector, (\ref{eq:Sec:eom-diff-full}). This will aid interpretation of the various terms.

We use the SVT split as defined in \cite{PhysRevD.76.023005}. Schematically, the components of the metric perturbation, $\ep g_{\mu\nu}$, perturbed energy-momentum tensor $\ep {T^{\mu}}_{\nu}$, and deformation vector $\xi^{\mu}$, are split as
\bse
\bea
\ep g_{\mu\nu} &\longrightarrow& \{ h, \eta, \qsuprm{H}{V}, \qsuprm{H}{T}\},\\
\ep {T^{\mu}}_{\nu}& \longrightarrow& \{ \delta\rho, \qsuprm{v}{S}, \delta P, \pis, \qsuprm{v}{V}, \qsuprm{\Pi}{V}, \qsuprm{\Pi}{T}\},\\
\xi^{\mu}& \longrightarrow&\{\qsuprm{\xi}{S},\qsuprm{\xi}{V}\}.
\eea
\ese
We will frequently use the density contrast, $\delta \defn \delta\rho/\rho$.

Under the SVT split, the perturbed gravitational field equations become \cite{PhysRevD.76.023005}
\bse
\label{eq:sec:svt-efeq-const}
\bea
\label{eq:sec:svt-3}
\hct\dot{h} - 2 k^2\eta &=& \kappa\delta\rho,\\
\label{eq:sec:svt-4}
2k\dot{\eta} &=& \kappa(\rho+P)\qsuprm{v}{S},\\
\label{eq:sec:svt-5}
k\dot{H}\qsuprm{}{V} &=& - 2\kappa(\rho+P)\qsuprm{v}{V},
\eea
\ese
\bse
\label{eq:sec:svt-efeq-evol}
\bea
\label{eq:sec:svt-6}
\ddot{h} + 2\hct\dot{h} - 2 k^2\eta &=& - 3\kappa\delta P,\\
\label{eq:sec:svt-7}
\ddot{h} + 6\ddot{\eta} + 2 \hct(\dot{h}+6\dot{\eta}) - 2 k^2\eta &=& -2\kappa P\qsuprm{\Pi}{S},\\
\label{eq:sec:svt-8}
\ddot{H}\qsuprm{}{V} + 2 \hct\dot{H}\qsuprm{}{V} &=& \kappa P\qsuprm{\Pi}{V},\\
\label{eq:sec:svt-9}
\ddot{H}\qsuprm{}{T} + 2 \hct\dot{H}\qsuprm{}{T} + k^2\qsuprm{H}{T} &=& \kappa P\qsuprm{\Pi}{T}.
\eea
\ese
with $ {\kappa} \defn 8 \pi Ga^2$.
The set of equations (\ref{eq:sec:svt-efeq-const}) are constraint equations, and (\ref{eq:sec:svt-efeq-evol}) are evolution equations.

\textit{\textbf{Gravity sources: scalar}}
The scalar parts of the components (\ref{eq:sec:pert-compts-ep-visco}) are
\bse
\label{eq:sec:pert-fluid-vars-fhdkjfhdj}
\bea
\label{eq:sec:pert-fluid-vars-fhdkjfhdj-a}
\delta\rho = -(\rho+P)\big( k\qsuprm{\xi}{S} +  \tfrac{1}{2} h\big),
\eea
\bea
\label{eq:sec:pert-fluid-vars-fhdkjfhdj-b}
\qsuprm{v}{S} = \dot{\xi}\qsuprm{}{S},
\eea
\bea
\label{eq:sec:pert-fluid-vars-fhdkjfhdj-c}
\delta P =-  \beta\big( k\qsuprm{\xi}{S} + \tfrac{1}{2} h\big)-  \lambda\big( k\dot{\xi}\qsuprm{ }{S} + \tfrac{1}{2} \dot{h}+2\hct[ k\qsuprm{\xi}{S} + \tfrac{1}{2} h]\big),\nonumber\\
\eea
\bea
\label{eq:sec:pert-fluid-vars-fhdkjfhdj-d}
P\qsuprm{\Pi}{S} &=& 2 {\mu}{ } \big( k\qsuprm{\xi}{S} +  \tfrac{1}{2} h + 3\eta\big) \nonumber\\
&&+ 2 {\nu}{ } \big( k\dot{\xi}\qsuprm{}{S} +  \tfrac{1}{2} \dot{h} + 3\dot{\eta} + 2\hct[k\qsuprm{\xi}{S} + \tfrac{1}{2} h + 3\eta]\big) .\nonumber\\
\eea
\ese

\textit{\textbf{Gravity sources: vector}} The vector parts of the components (\ref{eq:sec:pert-compts-ep-visco}) are
\bse
\bea
\label{eq:sec:vec-a}
\qsuprm{v}{V} = \dot{\xi}\qsuprm{}{V},
\eea
\bea
\label{eq:sec:vec-b}
P\qsuprm{\Pi}{V} &=& 2\mu\big( k\qsuprm{\xi}{V} -\qsuprm{H}{V}\big)+2\nu\big( k\dot{\xi}\qsuprm{ }{V} -\dot{H}\qsuprm{}{V} + 2 \hct[k\qsuprm{\xi}{V} -\qsuprm{H}{V}]\big).\nonumber\\
\eea
\ese

\textit{\textbf{Gravity sources: tensor}}
The tensor part of the components (\ref{eq:sec:pert-compts-ep-visco}) is
\bea
\label{eq:Sec:tensor-vsico}
P\qsuprm{\Pi}{T} = -2\mu \qsuprm{H}{T} -2\nu\big(   \dot{H}\qsuprm{}{T} + 2 \hct\qsuprm{H}{T}\big).
\eea

\textit{\textbf{Equation of motion: scalar}}
The scalar part of the equation of motion  (\ref{eq:Sec:eom-diff-full}) is
\bse
\bea
\label{eq:sec:eom-scal}
&&(\rho+P)[\ddot{\xi}\qsuprm{}{S} + \hct\dot{\xi}\qsuprm{}{S}] - 3[\beta\hct + \lambda(\dot{\hct} + \hct^2)]\dot{\xi}\qsuprm{}{S}  \nonumber\\
&&\qquad  + k^2(\lambda + \tfrac{4}{3}\nu)(\dot{\xi}\qsuprm{}{S} + 2\hct\qsuprm{\xi}{S})\nonumber\\
&&\qquad\qquad+ k^2(\beta + \tfrac{4}{3}\mu)\qsuprm{\xi}{S} = (\rho+P)\qsuprm{S}{scalar}_{[h]}, 
\eea
where the  source due to scalar metric perturbations is
\bea
(\rho+P)\qsuprm{S}{scalar}_{[h]} &=& - \tfrac{1}{2} k(\beta + \tfrac{4}{3}\mu)h -4k\mu\eta \nonumber\\
&&- \tfrac{1}{2}k(\lambda + \tfrac{4}{3}\nu)(\dot{h} + 2\hct h) - 4 k \nu(\dot{\eta} + 2 \hct\eta).\nonumber\\
\eea
\ese

\textit{\textbf{Equation of motion: vector}} The vector part of the equation of motion  (\ref{eq:Sec:eom-diff-full}) is
\bse
\bea
\label{eq:Sec:vec=eom-a}
&&(\rho+P)[\ddot{\xi}\qsuprm{}{V} + \hct\dot{\xi}\qsuprm{}{V}] - 3[\beta\hct + \lambda(\dot{\hct} + \hct^2)]\dot{\xi}\qsuprm{}{V}\nonumber\\
&&\qquad   + k^2\nu(\dot{\xi}\qsuprm{}{V} + 2\hct\qsuprm{\xi}{V})+ k^2\mu\qsuprm{\xi}{V}=(\rho+P)\qsuprm{S}{vector}_{[h]}, \nonumber\\
\eea
where the source due to the vector parts of the metric perturbations is
\bea
(\rho+P)\qsuprm{S}{vector}_{[h]} = k\mu\qsuprm{H}{V} + k\nu(\dot{H}\qsuprm{}{V} + 2 \hct\qsuprm{H}{V}).
\eea
\ese

The most transparent set of equations we could use to uncover the underlying physical behavior of the medium are those for the tensor modes, since they are by far the simplest. Using (\ref{eq:Sec:tensor-vsico}) to replace the tensor source of (\ref{eq:sec:svt-9}) yields 
\bea
\label{eq:sec:evolve-ht-with-tensor-source}
&&\ddot{H}\qsuprm{}{T} + 2[ \hct+8\pi G a^2\nu]\dot{H}\qsuprm{}{T} \nonumber\\
&&\qquad + [k^2 +16 \pi G a^2(\mu+2\hct\nu)] \qsuprm{H}{T} = 0.
\eea
It should now be  clear that (\ref{eq:sec:evolve-ht-with-tensor-source}) is the equation of motion of a massive field, $\qsuprm{H}{T}$, with damping. The damping has two contributions, each with a    different   physical origin. First, there is the usual Hubble damping, but there is also a term controlled by the coefficient of shear viscosity, $\nu$.  The mass-term also has two contributions: the first is a Hubble-independent contribution from the coefficient of shear elasiticity, $\mu$. Secondly, the coefficient of shear viscosity comes in, but multiplied by the Hubble expansion. In a flat background, (\ref{eq:sec:evolve-ht-with-tensor-source}) becomes
\bea
\ddot{H}\qsuprm{}{T} +   16\pi G \nu\dot{H}\qsuprm{}{T} + [k^2 +16 \pi G \mu] \qsuprm{H}{T} = 0,
\eea
further elucidating the physical mechanisms at play. 

The second thing we want to illustrate is the non-adiabatic nature of the medium. A medium is adiabatic if the pressure perturbation is specified entirely by the density perturbation. Using (\ref{eq:sec:pert-fluid-vars-fhdkjfhdj-a}) to rewrite (\ref{eq:sec:pert-fluid-vars-fhdkjfhdj-c}), we find
\bea
\delta P = \bigg[ \frac{\beta }{\rho+P} - \frac{\lambda}{\rho+P}\hct(1+3\tfrac{\dd P}{\dd\rho})\bigg] \delta\rho + \frac{\lambda}{\rho+P}\dot{\delta\rho}.\nonumber\\
\eea
It is clear that the pressure perturbation is determined by the density perturbation, and the rate-of-change of the density perturbation. The non-adiabaticity is controlled by the coefficient of bulk viscosity, $\lambda$.
 
We can perform some simple manipulations to provide evolution equations for the vector sources. First, inserting (\ref{eq:sec:vec-a}) into (\ref{eq:sec:vec-b}) yields
\bea
\label{eq:Sec:Ppi-justxi-legt}
P\qsuprm{\Pi}{V} &=& 2k\qsuprm{\xi}{V}(\mu+2\hct\nu) - 2 \mu\qsuprm{H}{V} \nonumber\\
&&+ 2 \nu(k\qsuprm{v}{V} - \dot{H}\qsuprm{}{V} - 2 \hct\qsuprm{H}{V}). 
\eea
Furthermore, inserting (\ref{eq:sec:vec-a}) into (\ref{eq:Sec:vec=eom-a}) yields
\bea
\label{eq:Sec:Ppi-justxi-legt-2}
k\qsuprm{\xi}{V}(\mu+2\hct\nu) &=&\tfrac{1}{k} (\rho+P)\qsuprm{S}{vector}_{[h]} - \nu k\qsuprm{v}{V} + 3 \beta\hct\tfrac{1}{k} \qsuprm{v}{V} \nonumber\\
&&- \tfrac{1}{k} (\rho+P)(\dot{v}\qsuprm{}{V} + \hct\qsuprm{v}{V}).
\eea
Combining (\ref{eq:Sec:Ppi-justxi-legt}) and (\ref{eq:Sec:Ppi-justxi-legt-2}) yields
\bea
\label{eq:sec:eom-vv}
\dot{v}\qsuprm{}{V} = - \hct\bigg(1- 3  \frac{\dd P}{\dd\rho} \bigg) \qsuprm{v}{V}-\frac{1}{2}\frac{ wk}{1+w}\qsuprm{\Pi}{V} ,
\eea
which is the equation of motion for the vector part of the velocity field, sourced by the vector part of the anisotropic stress. We can obtain an evolution equation for the vector anisotropic stress by differentiating (\ref{eq:Sec:Ppi-justxi-legt}), which yields
\bea
\label{eq:sec:eom-dotp-1}
P\dot{\Pi}\qsuprm{}{V} + \dot{P}\qsuprm{\Pi}{V} &=& 2 \mu(k\qsuprm{v}{V} - \dot{H}\qsuprm{}{V}) + 2 \nu(k\dot{v}\qsuprm{}{V} - \ddot{H}\qsuprm{}{V} \nonumber\\
&&+ 2 \dot{\hct}[k\qsuprm{\xi}{V} - \qsuprm{H}{V}] + 2 \hct[k\qsuprm{v}{V} - \dot{H}\qsuprm{}{V}]).\nonumber\\
\eea
Using the   equation of motion (\ref{eq:sec:eom-vv}) to replace $\dot{v}\qsuprm{}{V}$, and the gravitional equations (\ref{eq:sec:svt-5}) and (\ref{eq:sec:svt-8}) to replace $\dot{H}\qsuprm{}{V}$ and $\ddot{H}\qsuprm{}{V}$ respectively, the equation (\ref{eq:sec:eom-dotp-1}) becomes
\bea
\label{eq:sec:vector-evol}
&&P\dot{\Pi}\qsuprm{}{V} + \bigg[\frac{\dot{P}}{P} - \frac{2\nu\dot{\hct}}{\mu+2\nu\hct} + \frac{\nu k}{\rho(1+w)}   \bigg]P\qsuprm{\Pi}{V} \nonumber\\
   && =2\bigg(\mu k  + \nu\hct +3\nu\hct\frac{\dd P}{\dd\rho}-\frac{2\nu^2\dot{\hct}k}{\mu+2\hct\nu}  \bigg)\qsuprm{v}{V}   +  {4\nu\dot{\hct}} \qsuprm{H}{V}\nonumber\\
&&  + 2\kappa\bigg\{2\bigg[\mu+\frac{2\nu^2\dot{\hct}}{\mu+2\hct\nu} \bigg] \sum_{\rm{i}}(\qsubrm{\rho}{i}+\qsubrm{P}{i})\qsuprm{v}{V}_{\rm{i}}/k -  \nu\sum_{\rm{i}}\qsubrm{P}{i}\qsuprm{\Pi}{V}_{\rm{i}}\bigg\}.\nonumber\\
\eea
The term on the last line, proportional to $2\kappa=16\pi Ga^2$, contains sums over all matter species, including the viscoelastic medium. These terms arose after using the gravitational equations (\ref{eq:sec:svt-5}) and (\ref{eq:sec:svt-8}) to replace $\dot{H}\qsuprm{}{V}$ and $\ddot{H}\qsuprm{}{V}$ respectively.
\subsection{Propagation speeds and damping coefficients}

The  propagation speeds of the scalar and vector modes of the deformation vector can be read off from (\ref{eq:sec:eom-scal}) and (\ref{eq:Sec:vec=eom-a}) as the coefficients of $k^2\qsuprm{\xi}{S}$ and $k^2\qsuprm{\xi}{V}$ respectively. The damping coefficients can also be isolated, as the coefficients of $k\dot{\xi}\qsuprm{}{S}$ and $k\dot{\xi}\qsuprm{ }{V}$ respectively. This process yields the (dimensionless) scalar and vector sound speeds,
\bse
\label{eq:sec:scalarvector-speed-xi}
\bea
\label{eq:sec:scalar-speed-xi}
\qsubrm{c}{s}^2 \defn \frac{\beta + \tfrac{4}{3}\mu}{\rho+P} + 2 \hct\qsubrm{d}{s}  /k,
\eea
\bea
\qsubrm{c}{v}^2 \defn \frac{\mu}{\rho+P} + 2 \hct\qsubrm{d}{v}/k,
\eea
\ese
\bse
and the (dimensionless) scalar and vector damping coefficients,
\bea
\qsubrm{d}{s}  \defn \frac{\lambda + \tfrac{4}{3}\nu}{\rho+P}k,
\eea
\bea
\qsubrm{d}{v}\defn \frac{\nu}{\rho+P}k.
\eea
\ese
The sound speeds  (\ref{eq:sec:scalarvector-speed-xi}) should be compared with those we derived in the non-relativistic case, (\ref{eq:class-prop-speeds}).
Note that the scalar and vector propagation speeds are related via
\bea
\label{eq:sec:relate-cv-cs}
\qsubrm{c}{s}^2 =\frac{4}{3} \qsubrm{c}{v}^2 + \frac{\beta + 2 \hct\lambda}{\rho+P},
\eea
and the scalar and vector damping coefficients are related via
\bea
\qsubrm{d}{s} = \frac{4}{3}\qsubrm{d}{v} + \frac{\lambda k}{\rho+P}.
\eea
The material properties of the medium can be found in terms of the sound speeds and damping coefficients via
\bse
\label{mat-ceoffc-from-cs-cv-ds-dv}
\bea
\beta &=& (\rho+P)\big[ \qsubrm{c}{s}^2 - \tfrac{4}{3}\qsubrm{c}{v}^2 - 2 \hct(\qsubrm{d}{s} - \tfrac{4}{3}\qsubrm{d}{v})/k\big],\\
\mu &=& (\rho+P)\big[\qsubrm{c}{v}^2 - 2 \hct \qsubrm{d}{v}/k\big],\\
\lambda &=& (\rho+P)\big[ \qsubrm{d}{s} - \tfrac{4}{3}\qsubrm{d}{v}\big]/k,\\
\nu &=& (\rho+P)\qsubrm{d}{v}/k.
\eea
\ese
Using these definitions and the condition  (\ref{eq:sec:dpdrho-fromdecoupchi}) one can also obtain
\bea
\label{eq:sec:relate-cv-cs-useful-3}
\frac{\dd P}{\dd\rho} = (\qsubrm{c}{s}^2 - \tfrac{4}{3}\qsubrm{c}{v}^2) + (\qsubrm{d}{s} - \tfrac{4}{3}\qsubrm{d}{v})\frac{\dot{\hct} - \hct^2}{k\hct}.
\eea
The relationship (\ref{eq:sec:relate-cv-cs-useful-3}) will be very useful, especially when obtaining the conditions under which material   properties are constant.
\section{Equations of state for perturbations}
\label{sec:eos-visco}
The viscoelastic model of dark energy we have constructed can be written in the form of equations of state for perturbations \cite{PearsonBattye:eos, Battye:2013ida}. These equations of state are two functions (the entropy and anisotropic stress) which enter into the perturbed fluid equations, and parameterize all freedom in any dark energy or modified gravity model. These two functions are equations of state when they can be written in terms of fluid and metric variables alone. Once these expressions are identified, the task of computing observational signatures becomes simple.

The scalar perturbed fluid equations, for a general fluid (which may have $\dot{w}\neq0$, entropy, and anisotropic stress) are given by \cite{Hu:1998kj}
\bse
\label{eq:sec:fluid-eqs-gen}
\bea
\label{eq:sec:fluid-eqs-gen-pfe-dot-delta}
\left(\frac{\delta}{1+w}\right)^{\cdot}  = - \bigg( k \qsuprm{v}{S} + \tfrac{1}{2} \dot{h}\bigg) - \frac{3 \hct}{1+w} w\Gamma,
\eea
\bea
\dot{v}\qsuprm{}{S} = - \hct\bigg(1-3\frac{\dd P}{\dd\rho}\bigg) \qsuprm{v}{S} +\frac{k}{1+w} \bigg[\frac{\dd P}{\dd\rho}\delta + w\Gamma - \tfrac{2}{3} w\qsuprm{\Pi}{S}\bigg],\nonumber\\
\eea
\ese
where 
\bea
\label{eq:sec:entropp-gagin-v-defn}
w\Gamma \defn \bigg( \frac{\delta P}{\delta\rho} - \frac{\dd P}{\dd\rho} \bigg) \delta
\eea
is the entropy perturbation. The   equations of state for perturbations     prescribes the entropy and anisotropic stress in terms of variables which are already evolved. Put another way, we want to  eliminate $\qsuprm{\xi}{S}$ from the entropy and   anisotropic stress. We note that the scalar fluid equations (\ref{eq:sec:fluid-eqs-gen}) are gauge invariant.

From the scalar perturbed fluid variables (\ref{eq:sec:pert-fluid-vars-fhdkjfhdj-a}) and (\ref{eq:sec:pert-fluid-vars-fhdkjfhdj-b}) we have
\bea
\label{eq:Sec:xis-xisdot-drho-h-v}
k\qsuprm{\xi}{S} = - \frac{\delta\rho}{\rho+P} - \half h,\qquad \dot{\xi}\qsuprm{}{S} = \qsuprm{v}{S}.
\eea
The expressions (\ref{eq:Sec:xis-xisdot-drho-h-v})  can   be used in the pressure perturbation (\ref{eq:sec:pert-fluid-vars-fhdkjfhdj-c}), which can then be used to compute the entropy (\ref{eq:sec:entropp-gagin-v-defn}),   as well as the scalar anisotropic stress (\ref{eq:sec:pert-fluid-vars-fhdkjfhdj-d}) whilst making use of (\ref{eq:sec:dpdrho-fromdecoupchi})    and  (\ref{eq:sec:relate-cv-cs-useful-3}). This process yields
\bse
\label{eq:visco-eos-result}
\bea
\label{gamma-result-fhkjvghkdfjgbg}
w\Gamma = (\qsubrm{d}{s} - \tfrac{4}{3}\qsubrm{d}{v})k^{-1}\bigg[ \tfrac{\hct^2 - \dot{\hct}}{\hct} \delta -(1+w) (k\qsuprm{v}{S} + \tfrac{1}{2} \dot{h}) \bigg],\nonumber\\
\eea
\bea
\label{eq:sec:aniso-eos}
w\pis &=&\tfrac{3}{2}\big( \tfrac{\dd P}{\dd\rho} -\qsubrm{c}{s}^2+[\qsubrm{d}{s} - \tfrac{4}{3}\qsubrm{d}{v}]\tfrac{\hct^2 - \dot{\hct}}{k\hct}\big)\nonumber\\
&&\times\bigg[ \delta - 3 (1+w)\eta -\frac{\qsubrm{d}{v}}{k\qsubrm{c}{v}^2}(1+w)(\tfrac{1}{2}\dot{h} + k\qsuprm{v}{S} + 3 \dot{\eta}  )  \bigg].\nonumber\\
\eea
\ese
The expressions (\ref{eq:visco-eos-result}) are   equations of state for perturbations; specifically, the entropy and scalar anisotropic stress perturbation for a viscoelastic medium. The prefactors may   look    cumbersome, but they   allow for an intuitive understanding of the freedom in the equations of state for perturbations. For example, if the vector damping coefficient switches off, $\qsubrm{d}{v}=0$,   the relevant equations of state which describes a medium with this given physical property can be obtained quickly. This allows us to place physical restrictions on the medium with ease. The measurement of the sound speeds and damping coefficients $\{ \qsubrm{c}{s}^2,\qsubrm{c}{v}^2, \qsubrm{d}{s}, \qsubrm{d}{v}\}$ will allow measurement of the material properties $\{\beta,\mu,\lambda,\nu\}$ via (\ref{mat-ceoffc-from-cs-cv-ds-dv}).

In terms of the material properties $\{\beta,\mu,\lambda,\nu\}$, the equations of state for perturbations are
\bse
\label{eos-phys}
\bea
w\Gamma = \frac{\lambda}{\rho+P}\bigg[ \tfrac{\hct^2 - \dot{\hct}}{\hct}\delta - (1+w)(k\qsuprm{v}{S} + \tfrac{1}{2}\dot{h})\bigg],
\eea
\bea
w\pis &=& - 2 \frac{\mu+2\hct\nu}{\rho+P}\bigg[ \delta -3 (1+w)\eta\bigg]\nonumber\\
&& \qquad\qquad+ \frac{2\nu}{\rho}\bigg[ k\qsuprm{v}{S} + \tfrac{1}{2}\dot{h} + 3 \dot{\eta}\bigg].
\eea
\ese

The expressions for the entropy (\ref{gamma-result-fhkjvghkdfjgbg}) and anisotropic stress  (\ref{eq:sec:aniso-eos}) are gauge invariant. To show this we must perform a gauge transformation and show that all gauge artifacts cancel out. We first recall the following transformation rules for the fluid and metric perturbations from the synchronous gauge to the conformal Newtonian gauge:
\bse
\label{eq:sec:Trans-rules-cn-syn}
\bea
\delta\rho &=& \widehat{\delta\rho} -\dot{\rho}\zeta,\\
\qsuprm{v}{S} &=& \hat{v}  - \zeta k,\\
\dot{h} &=& - 6(\dot{\Phi} + \Psi\hct) + 2 \big[ k^2 - 3 (\dot{\hct} - \hct^2)\big]\zeta,\\
\eta &=& \Phi + \hct\zeta,\\
\dot{\eta} &=& \dot{\Phi} + \hct\Psi + (\dot{\hct} - \hct^2)\zeta,
\eea
\ese
where the hatted variables are those in the conformal Newtonian gauge, and   $\zeta$   is the gauge artifact. The   building blocks of fields in the entropy and anisotropic stress are   gauge invariant; performing a gauge transformation from the synchronous to conformal Newtonian gauge reveals that
\bse
\bea
&&\tfrac{\hct^2 - \dot{\hct}}{\hct} \delta -(1+w) (k\qsuprm{v}{S} + \tfrac{1}{2} \dot{h}) \nonumber\\
&&\qquad= \tfrac{\hct^2 - \dot{\hct}}{\hct} \hat{\delta} - (1+w)\big(k\hat{v}  - 3 (\dot{\Phi} + \Psi\hct)\big),\nonumber\\
\eea
\bea
\delta - 3 (1+w)\eta &=& \hat{\delta} - 3 (1+w)\Phi,
\\
\label{eq:sec:builingblock2}
\tfrac{1}{2}\dot{h} + k\qsuprm{v}{S} + 3 \dot{\eta} &=& k\hat{v}.
\eea
\ese
All gauge artifacts have dropped out automatically. Interestingly, notice that (\ref{eq:sec:builingblock2}) is just the velocity field in the conformal Newtonian gauge.  Therefore, we conclude that the  equations of state for perturbations (\ref{eq:visco-eos-result}) are  gauge invariant.

In a non-expanding background, $\hct=0$, and the scalar fluid equations (\ref{eq:sec:fluid-eqs-gen}) can be  combined to yield a second order evolution equation for density perturbations, 
\bea
\label{eq:sec:ddot-delta-flat}
\ddot{\delta}  + k^2[w\delta + w\Gamma - \tfrac{2}{3}w\pis] = - \tfrac{1}{2}(1+w)\ddot{h}  .
\eea
On a flat background, the viscoelastic equations of state for perturbations (\ref{eq:visco-eos-result}) become
\bse
\label{eq:Sec:flat-eos-visco}
\bea
w\Gamma = (\qsubrm{d}{s} - \tfrac{4}{3}\qsubrm{d}{v})\dot{\delta}/k,
\eea
\bea
w\pis &=&\tfrac{3}{2}\big(w-\qsubrm{c}{s}^2  \big)\nonumber\\
&&\times\bigg[ \delta - 3 (1+w)\eta+\frac{\qsubrm{d}{v}}{k\qsubrm{c}{v}^2}(\dot{\delta}- 3(1+w) \dot{\eta}  )  \bigg]. \nonumber\\
\eea
\ese
So, putting the flat-space viscoelastic equations of state for perturbations (\ref{eq:Sec:flat-eos-visco}) into (\ref{eq:sec:ddot-delta-flat}) yields
\bea
\label{eq:Sec:ddot-delta-flat-visco-with-eos}
&&\ddot{\delta} + k\qsubrm{d}{s}\dot{\delta} + k^2\qsubrm{c}{s}^2\delta \nonumber\\
&& = - \tfrac{1}{2}(1+w)\ddot{h} +      3 k^2(1+w)(\qsubrm{c}{s}^2-w)\big[\eta+\tfrac{\qsubrm{d}{v}}{\qsubrm{c}{v}^2} \dot{\eta}\big].\nonumber\\
\eea
This shows us that density waves are damped, with damping magnitude $\qsubrm{d}{s}$, and propagate with speed $\qsubrm{c}{s}^2$. This vindicates  $\qsubrm{c}{s}^2$ as being a sound speed. A simple observation to make from (\ref{eq:Sec:ddot-delta-flat-visco-with-eos}) is that viscosity plays no role in the sound speed of the viscoelastic medium: it is coefficients of elasticity which generate the sound speed, whilst the coefficients of   viscosity \textit{only}  modify the damping of density waves. This should be compared with, e.g., the ``viscosity parameter'', $\qsubrm{c}{vis}^2$, introduced in \cite{0004-637X-506-2-485} to parameterize somewhat adhoc modifications to the perturbed fluid equations.

The important point to take away from the equations of state for perturbations (\ref{eq:visco-eos-result}) is that the theory has prescribed which gauge invariant combinations are used to construct $w\Gamma$ and $w\pis$.

\section{Time variation of the physical properties}  
\label{sec:timevariation}
Constraining free functions of time with cosmological data is very hard to do, and so it is useful to have a consistent parameterization in which all of the freedom is contained within constants. A priori, all material properties, sound speeds, and damping coefficients are functions of time. If any of these lose their time variation, or if it is prescribed in some way, the theory tells us what that means for  the time variation of the other parameters. This is seen most easily by the relationships (\ref{eq:sec:dpdrho-fromdecoupchi}) and (\ref{mat-ceoffc-from-cs-cv-ds-dv}). We shall see that our material model can   consistently have all its freedom parameterized by constants, but that comes with important consistency conditions which we can derive.

Clearly, a few choices are possible; we will elucidate   two cases which seem rather natural. In our first case we will take   the sound speeds and damping coefficients $\{\qsubrm{c}{s}^2, \qsubrm{c}{v}^2, \qsubrm{d}{s}, \qsubrm{d}{v}\}$ to be constant, and let the material properties $\{\beta,\lambda,\mu,\nu\}$ vary in time. The second case is exactly the opposite: the material properties are constant, and the sound speeds and damping coefficients are time varying.


As for the cosmological background, the equation of state parameter, $w = P/\rho$, is of paramount importance in dark energy cosmology, as is its possible time variation; parameterizations have been devised which aim to capture this possible variation for various models. Our model of a viscoelastic medium gives us the allowed time variation ``for free'' as we will   show. However, our focus is to show under what circumstances   $w$ becomes constant, and what subsequent conditions get placed on the relationship between the material properties.

Since $P = w\rho$ and $\dot{P}/\dot{\rho} = \dd P/\dd\rho$, without loss of generality one can obtain
\bea
\dot{w} = \frac{\dot{\rho}}{\rho} \bigg( \frac{\dd P}{\dd\rho}-w\bigg).
\eea
Using our   expression for $\dot{P}$ which came from requiring time diffeomorphism invariance, (\ref{eq:sec:evol-press-re-decoup_chi}), we find
\bea
\label{eq:Sec:dot-w}
\dot{w} = - \frac{3}{\rho} \bigg[ \beta\hct  + \lambda ({\dot{\hct}+\hct^2}){ }- \hct (1+w)w\rho\bigg].
\eea
This is an evolution equation for the equation of state parameter, $w$. The particular   combination of the material parameters, $\beta$ and $\lambda$, that   give $\dot{w}=0$  is 
\bea
\beta\hct  + \lambda ({\dot{\hct}+\hct^2}){ }=\hct (1+w)w\rho.
\eea
When $\dot{w}=0$, the relationship (\ref{eq:sec:relate-cv-cs-useful-3}) becomes the constraint
\bea
\label{eq:sec:conf-foihfdlhfdlfhld}
(\dot{\hct} - \hct^2)(\qsubrm{d}{s} - \tfrac{4}{3}\qsubrm{d}{v}) = (w-\qsubrm{c}{s}^2 + \tfrac{4}{3}\qsubrm{c}{v}^2)k\hct.
\eea

\begin{figure}[!t]
      \begin{center}
{\includegraphics[scale=0.4,angle=0]{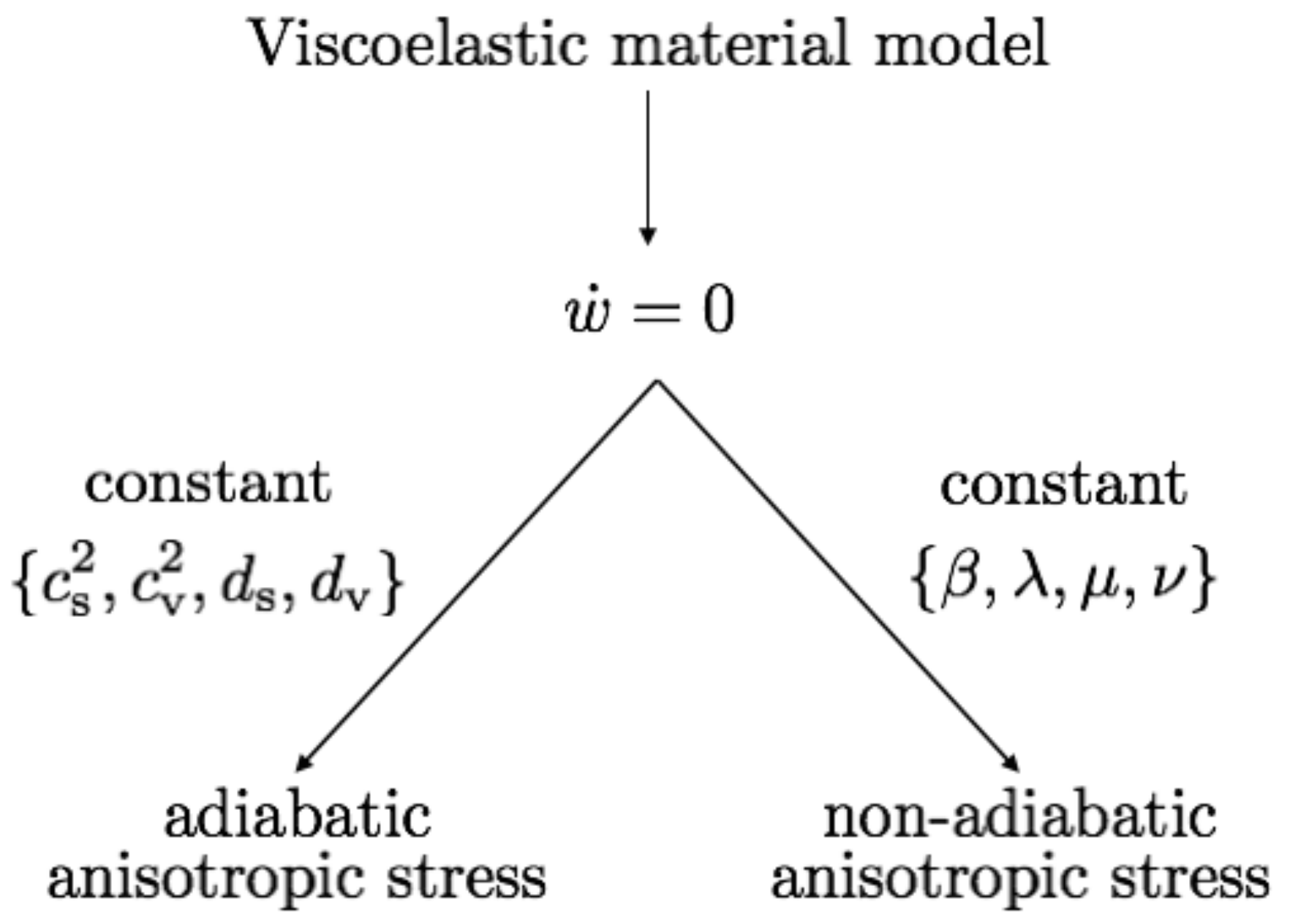}}
      \end{center}
\caption{\, Schematic illustration of what happens to the physics of the material when either the sound speeds and damping coefficients are constant, or the material properties are constant in the viscoelasic material model. In both cases there is anistropic stress, but forcing the sound speeds and damping coefficients to be constant means that the medium is adiabatic. These points are discussed in detail in Section \ref{sec:timevariation}.}\label{fig:shem_material}
\end{figure}

In what follows we will keep $w$ as constant, and then  take the sound speeds and damping coefficients to be constant, and then the material properties to be constant, and derive the susequent consistency conditions. The physical implications of this are summarised in  Figure \ref{fig:shem_material}.

\subsection{Constant sound speeds and damping coefficients}
We now proceed with the first of our special cases: when all of the sound speeds and damping coefficients are constant,
\bea
\dot{d}\qsubrm{}{s}=\dot{d}\qsubrm{}{v}=\dot{c}\qsubrm{}{s}^2=\dot{c}\qsubrm{}{v}^2=0,
\eea
the only way to satisfy the constraint (\ref{eq:sec:conf-foihfdlhfdlfhld}) for  arbitrary Hubble expansions $\hct$ is to set
\bse
\label{eq:conditions-wdot-constn-cond-soundspeed-damping}
\bea
\qsubrm{d}{s} - \tfrac{4}{3}\qsubrm{d}{v}  &=&0,\\
  w-\qsubrm{c}{s}^2 + \tfrac{4}{3}\qsubrm{c}{v}^2&=&0.
\eea
\ese
After imposing the constancy conditions (\ref{eq:conditions-wdot-constn-cond-soundspeed-damping}) and $\dot{w}=0$, the viscoelastic equations of state for perturbations (\ref{eq:visco-eos-result}) become
\bse
\bea
w\Gamma = 0,
\eea
\bea
\label{eq:constan-ve-pis}
w\pis &=&\tfrac{3}{2}\big(w-\qsubrm{c}{s}^2\big) \bigg[ \delta - 3 (1+w)\eta\bigg] \nonumber\\
&&+ \tfrac{3}{2}\qsubrm{d}{s}k^{-1}(1+w)\bigg[ k\qsuprm{v}{S} +3 \dot{\eta} +\tfrac{1}{2}\dot{h}     \bigg].
\eea
\ese
We now see that the medium is adiabatic, and there are three constants which parameterize the scalar perturbations of the viscoelastic medium: $\{ w, \qsubrm{c}{s}^2, \qsubrm{d}{s} \}$. We stress again that it is consistent to have these parameters being constant. It is interesting to note that the viscoelastic anisotropic stress (\ref{eq:constan-ve-pis}) can be written in terms of the   anisotropic stress for a perfectly elastic medium via
\bea
w {\Pi}^{{\rm \scriptscriptstyle{S}}}_{{\rm \scriptscriptstyle{vis}}} = w{\Pi}^{{\rm \scriptscriptstyle{S}}}_{{\rm \scriptscriptstyle{ela}}} + \frac{\qsubrm{d}{s}}{\qsubrm{c}{s}^2-w}k^{-1}w{\dot{\Pi}}{}^{{\rm \scriptscriptstyle{S}}}_{{\rm \scriptscriptstyle{ela}}} ,
\eea 
where
\bea
w{\Pi}^{{\rm \scriptscriptstyle{S}}}_{{\rm \scriptscriptstyle{ela}}} =\tfrac{3}{2}\big(w-\qsubrm{c}{s}^2\big) \bigg[ \delta - 3 (1+w)\eta\bigg] .
\eea
\subsection{Constant material properties}
Our second special case is where we allow the sound speeds and damping coefficients to be time dependent, but constrain the material properties $\{\beta, \mu,\lambda,\nu\}$ to be constant. From the condition (\ref{eq:sec:conf-foihfdlhfdlfhld}) we see that this is respected by the equations of state we presented in (\ref{eos-phys}), and repeat here for completeness:
\bse
\bea
w\Gamma = \frac{\lambda}{\rho+P}\bigg[ \tfrac{\hct^2 - \dot{\hct}}{\hct}\delta - (1+w)(k\qsuprm{v}{S} + \tfrac{1}{2}\dot{h})\bigg],
\eea
\bea
w\pis &=& - 2 \frac{\mu+2\hct\nu}{\rho+P}\bigg[ \delta -3 (1+w)\eta\bigg]\nonumber\\
&& \qquad\qquad+ \frac{2\nu}{\rho}\bigg[ k\qsuprm{v}{S} + \tfrac{1}{2}\dot{h} + 3 \dot{\eta}\bigg].
\eea
\ese
There are now four constants which parameterize the evolution of the medium, $\{w, \lambda, \mu,\nu\}$. The medium has retained its non-adiabaticity, and it is controlled by the coefficient of bulk viscosity.
 
\section{Viscosity and coupled dark energy theories}
\label{sec:connect}
We have constructed the gravitational field equations that describe a viscoelastic medium. An interesting question is to ask what types of gravitational theories, of a more conventional type, yield field equations which have a similar structure to those for a viscoelastic medium. In previous work, we uncovered a correspondance between the perfect elasticity theory and time diff. invariant massive gravity theories \cite{BattyePearson_connections}.

To uncover what type of gravity theory could yield something which looks like our viscoelastic theory, we will begin by understanding precisely how  the viscous terms modify the perfect elastic theory (which we already know corresponds to a Lorentz violating massive gravity theory). One of the obvious differences is that the perfect elastic theory could be constructed from a Lagrangian, but the viscous term cannot. This is a well known issue -- dissipative systems do not come from a Lagrangian. 

In this section, we will consider a theory which contains conventional matter, contributing an energy-momentum tensor $T^{\mu\nu}$ to the gravitational field equations, and a viscoelastic medium which is the sole contributor towards the dark energy-momentum tensor $U^{\mu\nu}$.

As   seen in the short review we gave in Section \ref{eq:sec:intro-rayleigh}, the equations of motion for dissipative systems can be constructed from a pair of invariants: the potential   and   Raleigh functions. Here we will write down   specific forms of    these invariants   for the relativistic viscoelastic solid we have been discussing. The energy-momentum tensor can be constructed from a pair of quadratic invariants, namely the Lagrangian for perturbations, $\sol$,  and  the generalized Rayleigh function, $\rol$ via
\bea
\label{eq:const-lp-U-sol-rol}
\lp U^{\mu\nu} = - \half\bigg[ 4\frac{\virt\sol}{\virt\lp g_{\mu\nu}} + 4\frac{\virt\rol}{\virt\lp K_{\mu\nu}} + U^{\mu\nu}g^{\alpha\beta}\lp g_{\alpha\beta}\bigg].\nonumber\\
\eea
The appropriate invariants that describe the relativistic viscoelastic medium are
\bse
\label{eq:sol-rol-visc}
\bea
\label{eq:sec:sol-pefalst}
\sol &=& \tfrac{1}{8} {W}^{\mu\nu\alpha\beta}\lp g_{\mu\nu}\lp g_{\alpha\beta},\\
\rol &=& \tfrac{1}{8} {V}^{\mu\nu\alpha\beta}\lp K_{\mu\nu}\lp K_{\alpha\beta}.
\eea
\ese
Using (\ref{eq:sol-rol-visc}) to compute (\ref{eq:const-lp-U-sol-rol}) gives precisely the energy-momentum tensor we presented in (\ref{eq:sec:lp-T}).  The tensor $W^{\mu\nu\alpha\beta}$ is exactly that which we defined in (\ref{eq:sec:gen-elast0tensor-fdkjfhd}). If we are using this formalism in which the Rayleigh function $\rol$ is used to compute the viscous contributions to $\lp U^{\mu\nu}$, then the viscosity tensor $V^{\mu\nu\alpha\beta}$ gains some more symmetries in its indices:
\bea
{V}^{\mu\nu\alpha\beta} = {V}^{(\mu\nu)(\alpha\beta)} ={V}^{\alpha\beta\mu\nu}.
\eea
This additional symmetry is only important for anisotropic media.

It is not possible to redefine $\sol$ to incorporate the viscous contributions that are encoded in $\rol$, in order to be able to compute $\lp U^{\mu\nu}$ from the single quadratic invariant $\sol$ in the conventional manner, namely via
\bea
\label{eq:sec:lp-u-conventional}
\lp U^{\mu\nu} = - \half\bigg[ 4\frac{\virt\sol}{\virt\lp g_{\mu\nu}} + U^{\mu\nu}g^{\alpha\beta}\lp g_{\alpha\beta}\bigg].
\eea
That said, we can ``reverse engineer'' the action for perturbations that gives the required field equations after using the variational principle. To see this, suppose that we have an action for perturbations  given by
\bea
\label{eq:sec:S2-rpotot}
\qsubrm{S}{\{2\}} &=& \int \dd^4x\,\sqrt{-g}\, \bigg[ (\tfrac{1}{8\pi G}\ep G^{\mu\nu} - \ep T^{\mu\nu}-\ep U^{\mu\nu})\ep g_{\mu\nu} \nonumber\\
&&\qquad\qquad+ 2 \xi_{(\mu}(\ep\nabla_{\nu)}U^{\mu\nu} -\ep F^{\mu})\bigg],
\eea
up to the addition of total derivatives.
The variational derivatives of the action $\qsubrm{S}{\{2\}}$ with respect to the fields    $\ep g_{\mu\nu}$ and   $\xi^{\mu}$ yields
\bse
\label{eq:sec:var-deroves-fdhfkdhk}
\bea
\frac{\virt}{\virt \ep g_{\mu\nu}} \qsubrm{S}{\{2\}}&=&\ep G^{\mu\nu} -8\pi G[ \ep T^{\mu\nu}+\ep U^{\mu\nu}],
\\
\frac{\virt}{\virt \xi_{\nu}} \qsubrm{S}{\{2\}} &=& \ep (\nabla_{\mu}U^{\mu\nu}) - \ep F^{\nu}.
\eea
\ese
Demanding the vanishing of the variational derivatives (\ref{eq:sec:var-deroves-fdhfkdhk}) in accord with the principle of least action yields the field equations
\bse
\bea
\ep G^{\mu\nu} &=&8\pi G[ \ep T^{\mu\nu}+\ep U^{\mu\nu}],\\
\label{eq:Sec:field-eq-from-action-ep-nab-f}
\ep (\nabla_{\mu}U^{\mu\nu}) &=&\ep F^{\nu}.
\eea
\ese
These are perturbed gravitational field equations, where the dark energy-momentum tensor satisifies a perturbed \textit{sourced} conservation equation. These sources (or, one can think of them as being forces) are due to a coupling in the action between the fields that constructed $\ep F^{\nu}$, and the $\xi^{\mu}$-field: it was the $\xi_{\mu}\ep F^{\mu}$-term.  

We can now use this way of thinking to isolate the term in the action which gives rise to the viscoelastic behavior. We will start from the Lagrangian for perturbations that gives the elasticity theory, (\ref{eq:sec:sol-pefalst}). 
The dark energy-momentum tensor is constructed from the single quadratic invariant $\sol$ given by (\ref{eq:sec:sol-pefalst})  using    the conventional expression (\ref{eq:sec:lp-u-conventional}) and yields
\bea
\lp U^{\mu\nu} = - \tfrac{1}{2}\big[  {W}^{\mu\nu\alpha\beta} + U^{\mu\nu}g^{\alpha\beta}\big] \lp g_{\alpha\beta} .
\eea
In the case of perfect elasticity the energy-momentum tensor satisfies the conservation equation, $\ep (\nabla_{\mu}U^{\mu\nu})=0$. In anticipation, we   modify the conservation equation to include the influence of a force,
\bea
\label{eq:sec:per-emt-with-force}
\ep (\nabla_{\mu}U^{\mu\nu}) = \ep F^{\nu}.
\eea
If we want the field equation (\ref{eq:sec:per-emt-with-force})  to be identical to that for the viscoelastic medium, namely (\ref{eq:sec:eom-t-ep}) with (\ref{eq:sec:ep-t-mixed-cov}) for $\ep {T^{\mu}}_{\nu}$, then we require   the force term  to be   given by
\bea
\label{eq:sec:things-epF-ray-defn}
\ep F^{\nu}  = \tfrac{1}{2}\nabla_{\mu}( {V}^{\mu\nu\alpha\beta}\lp K_{\alpha\beta}).
\eea
Putting (\ref{eq:sec:things-epF-ray-defn}) for $\ep F^{\mu}$ into the last term of the action (\ref{eq:sec:S2-rpotot}) and integrating by parts, yields
\bea
\qsubrm{S}{\{2\}} \supset\qsubrm{S}{\{2\}}^{\rm{[visc]}}&=& \int\dd^4x\sqrt{-g}\bigg[    {V}^{ \rho\sigma\mu\nu}\lp K_{\mu\nu}\nabla_{\rho}\xi_{\sigma}\bigg],\nonumber\\
\eea
and using (\ref{eq:Sec:lp-g-e-g-xi}), this can be written as
\bea
\qsubrm{S}{\{2\}}^{\rm{[visc]}}&=& \half\int\dd^4x\sqrt{-g}\bigg[    {V}^{ \rho\sigma\mu\nu}\lp K_{\mu\nu}(\lp g_{\rho\sigma} - \ep g_{ \rho\sigma})\bigg].\nonumber\\
\eea
The second manipulation highlights the fact that the ``extra term'' is related to the difference between two metric perturbations.
We could use (\ref{eq:sec:var-lp--ep-k-xi-dsgkhdsgfmdb}) to replace $\lp K_{\mu\nu}$ with $ \ep K_{\mu\nu}$ and the appropriate derivatives of $\xi_{\mu}$. Putting these peices together, the action for perturbations that yields the viscoelastic theory is
\bea
\qsubrm{S}{\{2\}}  &=& \int \dd^4x\, \sqrt{-g} \bigg[ \Diamond^2R +16\pi G\Diamond^2\qsubrm{\ld}{m} \nonumber\\
&&- \tfrac{1}{4}W^{\mu\nu\alpha\beta} \lp g_{\mu\nu} \lp g_{\alpha\beta} + {V}^{ \rho\sigma\mu\nu}\lp K_{\mu\nu}\nabla_{\rho}\xi_{\sigma}\bigg],\nonumber\\
\eea
where $\Diamond^nX \defn \tfrac{1}{\sqrt{-g}}\delta^n(\sqrt{-g}X)$ is the measure-weighted variation operator.

The outcome of this is that we can think of the viscoelastic theory in two ways:
\begin{enumerate}
\item The first is as a prescription of an energy-momentum tensor, which requires the pair of quadratic invariants: the Lagrangian for perturbations $\sol$ and the Rayleigh function $\rol$.
\item The second, is to think of the theory as a coupled massive gravity theory: the viscous term acts like a dissipative force on the right-hand-side of the conservation equation. 
\end{enumerate}

This line of reasoning   is useful as a tool-box for constructing consistent generalizations and modifications of the viscous field equations; although, it must be said that one would lose the neat physical interpretation of the theory describing a material.
Only using the metric, the most general force will be constructable from an expression of the form
\bea
\ep F^{\mu} &=& \qsubrm{C}{(0)}^{\mu\alpha\beta} \lp g_{\alpha\beta} +  \qsubrm{C}{(1)}^{\mu\lambda\alpha\beta}\nabla_{\lambda} \lp g_{\alpha\beta} \nonumber\\
&&\qquad +  \qsubrm{C}{(2)}^{\mu\lambda\sigma\alpha\beta} \nabla_{\lambda}\nabla_{\sigma}\lp g_{\alpha\beta} + \cdots.
\eea
The tensors $ \qsubrm{C}{(n)}^{\mu\alpha\beta\cdots}$ contain all freedom in the theory: the number of components of these tensors prescribes the number of free parameter or functions needed to characterise the theory.
Truncating to  the first  three terms   above, the   term in the Lagrangian for perturbations which will yield the source to the perturbed conservation equation is
\bea
\xi_{\mu}\ep F^{\mu}
&=& \qsubrm{C}{(0)}^{\mu\alpha\beta}\xi_{\mu} \lp g_{\alpha\beta} +  \qsubrm{C}{(1)}^{\mu\lambda\alpha\beta}\xi_{\mu}\nabla_{\lambda} \lp g_{\alpha\beta} \nonumber\\
&&   +  \qsubrm{C}{(2)}^{\mu\lambda\sigma\alpha\beta} \xi_{\mu}\nabla_{\lambda}\nabla_{\sigma}\lp g_{\alpha\beta} \nonumber\\
&=& (\qsubrm{C}{(0)}^{\mu\alpha\beta}  - \nabla_{\lambda}\qsubrm{C}{(1)}^{\mu\lambda\alpha\beta})\xi_{\mu}\lp g_{\alpha\beta}\nonumber\\
&&  - \qsubrm{C}{(1)}^{\mu\lambda\alpha\beta}\nabla_{\lambda}\xi_{\mu}\lp g_{\alpha\beta} \nonumber\\
&&   - (\nabla_{\lambda}\qsubrm{C}{(2)}^{\mu\lambda\sigma\alpha\beta}) \xi_{\mu}\nabla_{\sigma}\lp g_{\alpha\beta} \nonumber\\
&&  - \qsubrm{C}{(2)}^{\mu\lambda\sigma\alpha\beta} \nabla_{\lambda}\xi_{\mu}\nabla_{\sigma}\lp g_{\alpha\beta}.
\eea
To go between the equalities we integrated by parts.

\section{Discussion}
\label{sec:discussion}
In this article we reviewed, developed, and advocated \textit{material models of dark energy}. As should be clear from our presentation, these are rather different in nature from the conventional scalar field theories or modified gravity theories -- the material models are built in order to include the effects of a physical medium.

In the development of the material models, our main results are
\begin{itemize}
\item (\ref{eq:sec:lp-T}), the variation of the energy-momentum tensor,
\item (\ref{eq:visco-eos-result}), the viscoelastic equations of state for perturbations.
\end{itemize}
 
We have seen that it is natural for the medium to have constant equation of state parameter, $w$. This makes comparison against observational data much simpler than if $w$ were to be  time varying. We also saw that the medium can have constant sound speeds and damping coefficients, but that enforces adiabaticity of the medium (anisotropic stress is retained). On the other hand, if the material properties are constant instead of the sound speeds and damping coefficients, then the medium remains non-adiabatic, where the size of the entropy perturbation is controlled by the coefficient of bulk viscosity alone. These two cases are summarised in \fref{fig:shem_material}.

The use of a Rayleigh function may aid the systematic construction of coupled  dark energy models: it will certainly allow all of the freedom to be identified from all theories with given field content.
 
An interesting (and, depending on ones point of view, important) issue we have thus far shyed away from is the question of the \textit{nature} of the material  we are supposedly describing. That is: do we expect there to be some genuine viscoelastic solid pervading the Universe which is the direct cause of cosmic acceleration? If the answer is ``\textit{yes}'', then one can begin to ask questions about the micro-physical origin of the material. The idea of  ``frustated domain wall networks'' was pursued for some time \cite{Battye:1999eq, PhysRevD.60.043505, PhysRevD.76.023005, Avelino:2006xy, Battye:2010dk, Battye:2011ff}, but only as a single example of a possible realization of the medium (the idea somewhat relied on an observationally incompatible value of the dark energy equation of state parameter, $w$); of course, there may be some other set of structures in the Universe whose coarse grained dynamics are similar to a viscoelastic solid. If the answer is ``\textit{no}'', then the formalism developed here should be thought of as a useful gate-way  for importing relevant, consistent, and useful mathematical descriptions from solid-state physics into cosmology, in our example.  This, rather useful, agnostisism is rife in the implementation and useage of generalized descriptions of cosmological perturbations, in, for example, the ``PPF'' \cite{Baker:2011jy, Baker:2012zs}, ``EFT'' \cite{Bloomfield:2011wa, Bloomfield:2012ff, Bloomfield:2013efa}, and equations of state for perturbations \cite{Battye:2012eu, Battye:2013aaa} approaches. Each of these approaches may be employed in two ``modes'': (1) to describe the dynamics of perturbations for an explicitly given theory, and (2) as a prototype for the evolution of some unknown theory whose dynamics can be described by the particular flavour of the formalism which is written down. These two ``modes of use'' are something of an asset to these generalized descriptions, and are the analogue of the agnostisism outlined above. We do not offer a definitive opinion, and prefer to keep an open mind as to the possible ``reality'' of the material. 

The theoretical basis for material models of dark energy outlined here is only the beginning of a programme of research centering around these models. In addition to delving deeper into  uncovering issues of a more theoretical nature, such as the behavior of the medium in the strong-field regime,  we will need to ascertain the observational compatibility of the material models. For instance, one should note that the theory naturally prescribes new evolution rules for vector (\ref{eq:sec:vector-evol}) and tensor (\ref{eq:sec:evolve-ht-with-tensor-source}) modes, which should result in priors on the allowed values of the material properties. The way of writing all results as equations of state for perturbations (see Section \ref{sec:eos-visco}) makes this rather simple for implementation into numerical codes (such as {\tt CAMB} \cite{Lewis:1999bs}). As pointed out in the Introduction, we are preparing a paper which will contain the observational constraints on the sound speed for a perfectly elastic solid \cite{BattyePearsonMoss_edeconstraints}.

The nature of the dark energy, and the possibility of non-GR gravitational physics is one of the major open problems in modern cosmology.  The material models  developed in this article are a novel alternative to the ubiquitous scalar field models, and provide an almost unique way in which consistent modifications to  gravity   can be written down which (a) are parameterized by a small number of constants, and (b) have  direct physical interpretation.
 
\section*{Acknowledgements}
The author  is supported by the STFC Consolidated Grant ST/J000426/1, and would like to acknowledge Alex Barreira, Richard Battye,  Ruth Gregory,  Baojiu Li, Adam Moss, and Ian Moss for many interesting discussions, comments, and questions which led to improvements in this manuscript. The author is also   appreciative to the anonymous referees who have provided valuable comments on previous versions of the manuscript.
\appendix
\section{Variations}
\subsection{Variations and orthogonal tensors}
\label{sec:-ofsarof}
A covariant tensor is one with only lower indices, and an orthogonal tensor is one which has vanishing  contractions  with the time-like unit vector $u^{\mu}$ on any of its indices.  After decomposing the metric $g_{\mu\nu} = \gamma_{\mu\nu} - u_{\mu}u_{\nu}$, one can obtain the following  useful identities for variations:
\bse
\label{eq:sec:identities-useful}
\bea
\label{eq:sec:appenx-id-1}
\lp u^{\mu} &=& \tfrac{1}{2}u^{\mu}u^{\alpha}u^{\beta} \lp g_{\alpha\beta},
\eea
\bea
\label{eq:sec:appenx-id-2}
\lp u_{\mu} &=& ({\gamma^{\alpha}}_{\mu} - \tfrac{1}{2}u_{\mu}u^{\alpha})u^{\beta} \lp g_{\alpha\beta},
\eea
\bea
\label{eq:sec:appenx-id-3}
\lp \gamma_{\mu\nu} &=&  \lp g_{ \mu\nu} + 2 u_{(\mu} ({\gamma^{\alpha}}_{\nu)} - \tfrac{1}{2}u_{\nu)}u^{\alpha})u^{\beta} \lp g_{\alpha\beta},\nonumber\\
\eea
\bea
\lp g^{ \mu\nu} &=& - g^{\mu\alpha}g^{\beta\nu}\lp g_{\alpha\beta},
\eea
\bea
\label{eq:sec:appenx-id-5}
\lp \cs{\alpha}{\mu}{\nu} &=& g^{\alpha\beta}(\nabla_{(\mu}\lp g_{\nu)\beta} - \tfrac{1}{2}\nabla_{\beta}\lp g_{\mu\nu}).
\eea
Another useful identity is
\bea
\label{eq:sec:identity-deri-gamma}
\pd{\gamma_{\mu\nu}}{\gamma_{\alpha\beta}} = {\gamma^{(\alpha}}_{\mu}{\gamma^{\beta)}}_{\nu}.
\eea
\ese
One should be careful when raising and lowering indices; for any symmetric 2-tensor $A^{\mu\nu}$, the perturbed contravariant components are related to the perturbed covariant components via
\bea
\lp A^{\mu\nu} &=& g^{\mu\alpha}g^{\beta\nu} \lp A_{\alpha\beta} - 2A^{\alpha(\mu}g^{\nu)\beta} \lp g_{\alpha\beta}.
\eea
For any quantity constructed as
\bea
\lp { B}_{\mu\nu} = \pd{{  B}_{\mu\nu} }{\gamma_{\alpha\beta}} \lp \gamma_{\alpha\beta},
\eea
the identity (\ref{eq:sec:appenx-id-3}) can be used  to replace the variation of the strain $ \lp \gamma_{\alpha\beta}$, giving
\bea
\lp B_{\mu\nu} = \bigg[ \pd{B_{\mu\nu}}{\gamma_{\alpha\beta}} + 2 \pd{B_{\mu\nu}}{\gamma_{\rho\sigma}}u_{(\rho}({\gamma^{\alpha}}_{\sigma)} - \tfrac{1}{2}u_{\sigma)}u^{\alpha})u^{\beta}\bigg] \lp g_{\alpha\beta}.\nonumber\\
\eea
Note that in the case where $ {\partial B_{\mu\nu}}/{\partial\gamma_{\rho\sigma}}$ is orthogonal on all indices, the ``complicated'' term above vanishes.
For a symmetric orthogonal tensor $C_{\mu\nu}$,
\bea
\label{eq:sec:show-raise-ortho}
\pd{C_{\mu\nu}}{\gamma_{\alpha\beta}} = \pd{(\gamma_{\mu\rho}\gamma_{\nu\sigma}C^{\rho\sigma})}{\gamma_{\alpha\beta}}  = \gamma_{\mu\rho}\gamma_{\nu\sigma}\pd{C^{\rho\sigma}}{\gamma_{\alpha\beta}} + 2 {C^{(\alpha}}_{(\mu}{\gamma^{\beta)}}_{\nu)}.\nonumber\\
\eea
This will be useful in computing the contravariant components of orthogonal tensors from the covariant ones.

If $X_{\mu\nu}$ is a covariant orthogonal tensor function of strain, then
\bea
\label{eq:sec:lp-P-lower}
\lp {  X}_{\mu\nu} = \pd{{  X}_{\mu\nu} }{\gamma_{\alpha\beta}} \lp \gamma_{\alpha\beta} =  \pd{{  X}_{\mu\nu} }{\gamma_{\alpha\beta}}\lp g_{\alpha\beta},
\eea
where the fact that term $ {\partial{  X}_{\mu\nu} }/{\partial\gamma_{\alpha\beta}}$ is orthogonal on all indices has been used after (\ref{eq:sec:appenx-id-3}) was inserted. Using (\ref{eq:sec:show-raise-ortho}) and (\ref{eq:sec:lp-P-lower}), we see that the contravariant components of the variation of an orthogonal tensor function of strain are given by
\bea
\lp X^{\mu\nu} = \bigg[ \pd{X^{\mu\nu}}{\gamma_{\alpha\beta}} +2X^{\alpha(\mu}u^{\nu)}u^{\beta} \bigg] \lp g_{\alpha\beta}.
\eea
If $Y_{\mu\nu}$ is an orthogonal covariant tensor function of strain \textit{and} rate of strain, then
\bea
\label{eq:sec:Lp-P-proto}
\lp Y_{\mu\nu} = \pd{Y_{\mu\nu}}{\gamma_{\alpha\beta}} \lp \gamma_{\alpha\beta} + \pd{Y_{\mu\nu}}{\lambda_{\alpha\beta}} \lp \lambda_{\alpha\beta}.
\eea
In what follows we will be concentrating on computing $\lp Y^{\mu\nu}$, and learning how to replace the variation of the rate of strain with space-time fields.

The variation of the rate of strain tensor is given by
\bea
\lp \lambda_{\mu\nu} = 2 \lp K_{\mu\nu} &=& 2\nabla_{\mu}\lp u_{\nu} - 2 u_{\alpha}\lp \cs{\alpha}{\mu}{\nu}.
\eea
And so it follows that
\bea
\pd{Y_{\mu\nu}}{\lambda_{\alpha\beta}}\lp \lambda_{\alpha\beta} = \pd{Y_{\mu\nu}}{K_{\alpha\beta}} \lp K_{\alpha\beta} ,
\eea
meaning that  (\ref{eq:sec:Lp-P-proto}) becomes
\bea
\label{eq:sec:lp-P-fljdhjkfhd}
\lp Y_{\mu\nu} &=&  \pd{Y_{\mu\nu}}{\gamma_{\rho\sigma }}  \lp g_{\rho\sigma}  +  \pd{Y_{\mu\nu}}{K_{\alpha\beta}}  \lp K_{\alpha\beta}.
\eea
Since  $Y_{\mu\nu}$ is an orthogonal tensor, using the identity (\ref{eq:sec:identity-deri-gamma}) we obtain
\bse
\label{eq:sec;change-upp-todwnn}
\bea
\pd{Y_{\mu\nu}}{\gamma_{\alpha\beta}} =\gamma_{\mu\rho}\gamma_{\nu\sigma} \pd{Y^{\rho\sigma}}{\gamma_{\alpha\beta}}  + 2{\gamma^{(\alpha}}_{(\mu}{Y^{\beta)}}_{\nu)},
\eea
\bea
\label{eq:sec;change-upp-todwnn-2}
\pd{Y_{\mu\nu}}{\lambda_{\alpha\beta}} = \gamma_{\mu\rho}\gamma_{\nu\sigma}\pd{Y^{\rho\sigma}}{\lambda_{\alpha\beta}}.
\eea
\ese
The expression (\ref{eq:sec:lp-P-fljdhjkfhd}) will prove very useful.

\subsection{Variation of the extrinsic curvature tensor}
\label{sec:var-extr-curv}
We will now show how to compute $\lp K_{\alpha\beta}$ in terms of its Eulerian perturbation, $\ep K_{\alpha\beta}$, and the corresponding contributions due to the deformation vector $\xi^{\mu}$. Since by definition $K_{\mu\nu} = \nabla_{\mu}u_{\nu}$, the components of the Lagrangian perturbed extrinsic curvature tensor are given by
\bea
\label{eq:sec:lp-K-proto}
\lp K_{\mu\nu} = \nabla_{(\mu}\lp u_{\nu)} - u_{\alpha}\lp\cs{\alpha}{\mu}{\nu}.
\eea
Using (\ref{eq:sec:appenx-id-2}) and (\ref{eq:sec:appenx-id-5}) for the components of the perturbed time-like vector and Christoffel symbols in   (\ref{eq:sec:lp-K-proto}) yields
\bea
\label{eq:sec:var-ext-tensor-general-eos}
\lp K_{\mu\nu} &=& \tfrac{1}{2}[K_{\mu\nu}u^{\alpha}u^{\beta} + 2{\gamma^{\alpha}}_{\mu}{K^{\beta}}_{\nu}]\lp g_{\alpha\beta} \nonumber\\
&&+ \tfrac{1}{2}[u^{\alpha}u^{\beta}u_{(\mu}{\gamma^{\sigma}}_{\nu)} + {\gamma^{\alpha}}_{\mu}{\gamma^{\beta}}_{\nu}u^{\sigma} \nonumber\\
&&- 2 {\gamma^{(\alpha}}_{\mu}u^{\beta)}u_{\nu}u^{\sigma}]\nabla_{\sigma}\lp g_{\alpha\beta}.
\eea
The projections of (\ref{eq:sec:var-ext-tensor-general-eos}) are
\bse
\bea
2{\gamma^{\mu}}_{\lambda}{\gamma^{\nu}}_{\pi} \lp K_{\mu\nu}  &=& {\gamma^{\alpha}}_{\lambda}{\gamma^{\beta}}_{\pi}u^{\sigma} \nabla_{\sigma}\lp g_{ \alpha\beta}\nonumber\\
&&+   \big[K_{ \lambda\pi}u^{\alpha}u^{\beta}  +2{\gamma^{\alpha}}_{\lambda}  {K^{\beta}}_{\pi}  \big]\lp g_{\alpha\beta} \nonumber\\
&=&  {\gamma^{\alpha}}_{\lambda}{\gamma^{\beta}}_{\pi}\lied{u}\lp g_{\alpha\beta} + K_{\lambda\pi}u^{\alpha}u^{\beta} \lp g_{\alpha\beta},\nonumber\\
\eea
\bea
-2u^{\nu}\lp K_{\mu\nu} &=& \big[  \tfrac{1}{2}  u^{\alpha}u^{\beta}{\gamma^{\sigma}}_{\mu} -2{\gamma^{(\alpha}}_{\mu}u^{\beta)}u^{\sigma}  \big]\nabla_{\sigma}\lp g_{ \alpha\beta}\nonumber\\
&=& \tfrac{1}{2}{\gamma^{\sigma}}_{\mu}\nabla_{\sigma}(u^{\alpha}u^{\beta} \lp g_{\alpha\beta})  - {K^{(\alpha}}_{\mu}u^{\beta)} \lp g_{\alpha\beta}\nonumber\\
&&- 2 u^{\lambda}{\gamma^{(\alpha}}_{\mu}\nabla_{\lambda}(u^{\beta)} \lp g_{\alpha\beta}),
\eea
\bea
u^{\mu}u^{\nu}\lp K_{\mu\nu}=0.
\eea
\ese
Replacing the Lagrangian variation of the metric with the Eulerian variation and corresponding Lie derivative, (\ref{eq:Sec:lp-g-e-g-xi}) 
elucidates   all contributions in the Lagrangian perturbed extrinsic curvature tensor (\ref{eq:sec:var-ext-tensor-general-eos}) due to the deformation field:
\bea
\lp K_{\mu\nu} &=&\tfrac{1}{2}(K_{\mu\nu}u^{\alpha}u^{\beta} + 2{\gamma^{\alpha}}_{\mu}{K^{\beta}}_{\nu})[\ep g_{\alpha\beta}+2\nabla_{(\alpha}\xi_{\beta)}] \nonumber\\
&&+ \tfrac{1}{2}{\mathfrak{k}^{\sigma\alpha\beta}}_{\mu\nu}[\nabla_{\sigma}\ep g_{\alpha\beta}  +2\nabla_{\sigma}\nabla_{(\alpha}\xi_{\beta)}], 
\eea
where we defined, for convience, the tensor
\bea
{\mathfrak{k}^{\sigma\alpha\beta}}_{\mu\nu}\defn (u^{\alpha}u^{\beta}u_{(\mu}{\gamma^{\sigma}}_{\nu)} + {\gamma^{\alpha}}_{\mu}{\gamma^{\beta}}_{\nu}u^{\sigma} - 2 {\gamma^{(\alpha}}_{\mu}u^{\beta)}u_{\nu}u^{\sigma}).\nonumber\\
\eea
It is also enlightening to write the Lagrangian perturbations $\lp u_{\mu}$ and $\lp \cs{\alpha}{\mu}{\nu}$ in terms of their Eulerian perturbations and the contribution due to the deformation field. After using (\ref{eq:Sec:lp-g-e-g-xi}) in (\ref{eq:sec:appenx-id-2}) and (\ref{eq:sec:appenx-id-5}) we find
\bse
\label{eq:sec:lp-u-ep-u-lp-cs-ep=cs}
\bea
\lp u_{\mu} &=& \ep u_{\mu} \nonumber\\
&&+ 2 ({\gamma^{\alpha}}_{\mu} - \tfrac{1}{2}u_{\mu}u^{\alpha})u^{\beta} \nabla_{(\alpha}\xi_{\beta)},\\
\lp \cs{\alpha}{\mu}{\nu} &=& \ep \cs{\alpha}{\mu}{\nu} \nonumber\\
&&+  \nabla_{(\mu}\nabla_{\nu)}\xi^{\alpha} + {R^{\alpha}}_{(\mu\nu)\beta}\xi^{\beta},
\eea
\ese
where the Eulerian perturbations are given by the usual expressions,
\bse
\bea
\ep u_{\mu} &=& ({\gamma^{\alpha}}_{\mu} - \tfrac{1}{2}u_{\mu}u^{\alpha})u^{\beta} \ep g_{\alpha\beta},\\
\ep \cs{\alpha}{\mu}{\nu} &=& g^{\alpha\beta}(\nabla_{(\mu}\ep g_{\nu)\beta} - \tfrac{1}{2}\nabla_{\beta}\ep g_{\mu\nu}),
\eea
\ese
and the background Riemann tensor is defined as
\bea
(\nabla_{\nu}\nabla_{\beta} - \nabla_{\beta}\nabla_{\nu})\xi_{\mu} = R_{\beta\mu\nu\alpha}\xi^{\alpha}.
\eea
Hence, using (\ref{eq:sec:lp-u-ep-u-lp-cs-ep=cs}) in (\ref{eq:sec:lp-K-proto}) gives the desired expression, namely
\bea
\label{eq:sec:var-lp--ep-k-xi-dsgkhdsgfmdb}
\lp K_{\mu\nu}  &=& \ep K_{\mu\nu}+2u^{(\alpha} {\gamma^{\beta)}}_{(\mu}\nabla_{\nu)} \nabla_{\alpha}\xi_{\beta} \nonumber\\
&&-  u^{\alpha}u^{\beta}u_{(\mu}\nabla_{\nu)} \nabla_{\alpha}\xi_{\beta}-  u^{\alpha}\nabla_{(\mu}\nabla_{\nu)}\xi_{\alpha} \nonumber\\
&&  + 2\big[({\gamma^{\alpha}}_{(\mu} - \tfrac{1}{2}u_{(\mu}u^{\alpha}){K^{\beta}}_{\nu)} + u^{\beta}{K^{\alpha}}_{(\mu}u_{\nu)}\big]\nabla_{(\alpha}\xi_{\beta)}  \nonumber\\
&&- u_{\alpha}{R^{\alpha}}_{(\mu\nu)\beta}\xi^{\beta}.
\eea
Note that the third and fourth line drops out on a flat background (since there, $K_{\mu\nu}=0$ and the Riemann tensor vanishes). We   gave explicit expressions for the components of (\ref{eq:sec:lp-k-withxiexpliciy}) in (\ref{eq:sec:comps-lagper-extcurv-syn}).
\providecommand{\href}[2]{#2}\begingroup\raggedright\endgroup

\end{document}